\documentclass{article}

\usepackage{amssymb}
\usepackage{stmaryrd}
\usepackage{amsmath}

\usepackage{cmll}
\usepackage{rotating}
\renewcommand{\parr}{\mathrel{\raisebox{1.7ex}{\begin{turn}{180}\&\end{turn}}}}

\usepackage{proof}
\usepackage[all]{xy}

\usepackage{graphicx}
\usepackage{color}

\newtheorem{thm}{Theorem}
\newtheorem{prop}[thm]{Proposition}
\newtheorem{lem}[thm]{Lemma}
\newtheorem{claim}{Claim}
\newtheorem{defn}{Definition}
\newenvironment{pf}{\textsc{Proof}. \nobreak}{}
\newcommand{\qed}{\hbox{}\hfill$\square$\smallbreak}

\newcommand{\scalefact}{1}

\newcommand{\refsect}[1]{Sect.~\ref{sect:#1}}

\newcommand{\refdef}[1]{Definition~\ref{def:#1}}
\newcommand{\refth}[1]{Theorem~\ref{th:#1}}
\newcommand{\refprop}[1]{Proposition~\ref{prop:#1}}
\newcommand{\reflemma}[1]{Lemma~\ref{lemma:#1}}

\newcommand{\refclaim}[1]{Claim~\ref{claim:#1}}
\newcommand{\reffig}[1]{Fig.~\ref{fig:#1}}
\newcommand{\reftab}[1]{Table~\ref{tab:#1}}

\def\cB{\mathcal B}
\def\cC{\mathcal C}
\def\cD{\mathcal D}

\def\lac{\mbox{$\lambda$-calculus}}
\def\lat{\mbox{$\lambda$-term}}

\def\pax{auxiliary port}
\def\pal{principal port}
\def\struct{discharged formula}

\newcommand{\Nat}{\mathbb N}
\newcommand{\Int}{\mathbb Z}
\newcommand{\iso}{\cong}
\newcommand{\act}{\cdot}

\def\MELL{$\mathbf{meLL}$}
\def\ELL{$\mathbf{ELL}$}
\def\LLL{$\mathbf{LLL}$}

\def\SLL{$\mathbf{SLL}$}
\def\DLAL{$\mathbf{DLAL}$}
\def\MLLL{$\mathbf{mLLL}$}
\def\MultELL{$\mathbf{mELL}$}
\def\MELLparg{$\mathbf{meLL}_{\S\mathrm{box}}$}
\def\MELLlev{$\mathbf{mL^3}$}
\def\MLLLlev{$\mathbf{mL^4}$}
\def\MELLz{$\mathbf{meLL_0}$}
\def\MLLLlevz{$\mathbf{mL_0^4}$}
\def\TwoELL{$\mathbf{2ELL}$}
\def\LLlev{$\mathbf{L^3}$}
\def\LLLlev{$\mathbf{L^4}$}

\newcommand{\ltens}{\otimes}

\newcommand{\lpar}{\parr}
\newcommand{\llinimp}{\multimap}
\newcommand{\lplus}{\oplus}
\newcommand{\lwith}{\with}
\newcommand{\parg}{\S}
\newcommand{\form}{\textit{Form}}
\newcommand{\canform}{\textit{Form}_0}
\newcommand{\cansubst}[3]{#1\{ #2 \slash #3 \} }
\newcommand{\ts}{\vdash}
\newcommand{\etq}[1]{\mbox{\scriptsize #1}}
\newcommand{\OneRule}[1]{
\begin{center}
	\makebox[\textwidth][c]{
		\makebox[\textwidth][c]{
			\begin{minipage}[c]{\textwidth}#1\end{minipage}
		}
	}
\end{center}
}
\newcommand{\TwoRules}[2]{
\begin{center}
	\makebox[\textwidth][c]{
		\makebox[0.5\textwidth][c]{
			\begin{minipage}[c]{0.5\textwidth}#1\end{minipage}
		}
		\makebox[0.5\textwidth][c]{
			\begin{minipage}[c]{0.5\textwidth}#2\end{minipage}
		}
	}
\end{center}
}

\def\pps{pre-net}
\def\ps{net}
\def\sps{sequentializable \ps}
\def\pn{proof net}
\def\Ps{Net}
\def\Sps{Sequentializable \ps}
\def\Pn{Proof net}

\def\axiom{$\mathsf{axiom}$}
\def\cut{$\mathsf{cut}$}
\def\tensor{$\mathsf{tensor}$}
\def\parl{$\mathsf{par}$}
\def\univ{$\mathsf{for\ all}$}
\def\exist{$\mathsf{exists}$}
\def\ofcourse{$\mathsf{of\ course}$}
\def\whynot{$\mathsf{why\ not}$}
\def\weak{$\mathsf{weakening}$}
\def\flatl{$\mathsf{flat}$}
\def\paxl{$\mathsf{pax}$}
\def\pargl{$\mathsf{paragraph}$}

\newcommand{\axlink}{\textsf{\footnotesize ax}}
\newcommand{\cutlink}{\textsf{\footnotesize cut}}
\newcommand{\paxlink}{\textsf{\footnotesize pax}}

\newcommand{\Size}[1]{|#1|}
\newcommand{\ISize}[2]{|#2|_{#1}}
\newcommand{\Depth}[1]{\mathop{\mathrm{d}}(#1)}
\newcommand{\RelDepth}[1]{\rho(#1)}
\newcommand{\Level}[1]{\ell(#1)}

\newcommand{\onered}{\rightarrow}
\newcommand{\red}{\onered^\ast}
\newcommand{\oneetared}{\rightarrow_\eta}
\newcommand{\etared}{\oneetared^\ast}

\newcommand{\boxprec}{\subseteq}
\newcommand{\boxprecneq}{\subset}

\newcommand{\ctrprec}{\preceq}
\newcommand{\immctrprec}{\prec_1}
\newcommand{\lghtctrprec}{\ctrprec^{\mathbf{L}}}
\newcommand{\immlghtctrprec}{\immctrprec^{\mathbf{L}}}
\newcommand{\MBox}[1]{\widehat{#1}}

\newcommand{\CtrFact}[1]{\mu(#1)}
\newcommand{\Arity}[1]{\nabla(#1)}
\newcommand{\Mult}[1]{\delta(#1)}
\newcommand{\PotSize}[1]{[#1]}
\newcommand{\Compl}{\sharp}
\newcommand{\Wght}[1]{\alpha_{#1}}
\newcommand{\Cuts}[2]{\mathsf{cuts}_{#1}(#2)}
\newcommand{\AllCuts}[1]{\mathsf{cuts}(#1)}
\newcommand{\Tree}[1]{\mathcal{T}(#1)}

\newcommand{\trzero}[1]{#1_{\mathbf{0}}}
\newcommand{\trone}[1]{#1_{\mathbf{1}}}
\newcommand{\Forget}{U}
\newcommand{\Erase}[1]{#1^-}
\newcommand{\NF}{\mathrm{NF}}

\newcommand{\Num}{\mathbf{N}}
\newcommand{\Str}{\mathbf{S}}
\newcommand{\ElStr}{\mathbf{S}_{\mathbf{E}}}
\newcommand{\PolyStr}{\mathbf{S}_{\mathbf{P}}}
\newcommand{\Strz}{\mathbf{S_0}}

\def\FE{$\mathbf{FE}$}
\def\FP{$\mathbf{FP}$}
\def\FMELLlev{$\mathbf{F{\mbox{\MELLlev}}}$}
\def\FMLLLlev{$\mathbf{F{\mbox{\MLLLlev}}}$}

\def\citep{\cite}

\title{Linear Logic by Levels\\and Bounded Time Complexity}
\author{
\begin{tabular}{cc}
	Patrick Baillot\footnote{\texttt{patrick.baillot@ens-lyon.fr}} & Damiano Mazza\footnote{\texttt{damiano.mazza@lipn.univ-paris13.fr}} \\
	\small ENS Lyon, Universit\'e de Lyon, LIP & \small CNRS-Universit\'e Paris 13, LIPN \\
	\footnotesize (UMR 5668 CNRS-ENSL-INRIA-UCBL) & \footnotesize (UMR 7030 CNRS-UP13)
\end{tabular}
}
\date{}

\begin{document}

\maketitle

\begin{abstract}
	\noindent We give a new characterization of elementary and deterministic polynomial time computation in linear logic through the proofs-as-programs correspondence. Girard's seminal results, concerning elementary and light linear logic, achieve this characterization by enforcing a \emph{stratification principle} on proofs, using the notion of \emph{depth} in proof nets. Here, we propose a more general form of stratification, based on inducing \emph{levels} in proof nets by means of indexes, which allows us to extend Girard's systems while keeping the same complexity properties. In particular, it turns out that Girard's systems can be recovered by forcing depth and level to coincide. A consequence of the higher flexibility of levels with respect to depth is the absence of boxes for handling the paragraph modality. We use this fact to propose a variant of our polytime system in which the paragraph modality is only allowed on atoms, and which may thus serve as a basis for developing lambda-calculus type assignment systems with more efficient typing algorithms than existing ones.
\end{abstract}

\section*{Introduction}
\paragraph*{Linear logic and implicit computational complexity.} The intersection between logic and implicit computational complexity is at least twofold, as there are at least two alternative views on logic itself: a first possibility is to see it as a \emph{descriptive language}, i.e., as a language for expressing properties of mathematical objects; a second possibility is to see it, via the Curry-Howard isomorphism, as a \emph{programming language}, i.e., a tool for computing functions. These two views closely correspond to two fundamental branches of mathematical logic: model theory, and proof theory, respectively. The first approach has been taken quite successfully by what is known as \emph{descriptive computational complexity}. The idea of exploring the second approach is more recent: the first results of this kind can be found in~\cite{Leivant94}, \cite{LeivantMarion93}, and in the work of \cite{GirardScedrovScott92}, to which the present work is more closely related.

As mentioned above, the use of logic as a programming language capturing certain complexity classes passes through the Curry-Howard isomorphism: a proof is a program, whose execution is given by cut-elimination; therefore, the idea is to define a logical system whose cut-elimination procedure has a bounded complexity, so that the algorithms programmable in this logical system intrinsically have that complexity, i.e., the system is \emph{sound} w.r.t.\ a complexity class.

\sloppy{Due to its ``resource awareness'', linear logic \citep{Girard:LL} is the ideal setting to attempt this. In fact, linear logic brings to light the logical primitives which are responsible for the complexity of cut-elimination, under the form of modalities, called \emph{exponentials}. These are in control of duplication during the cut-elimination process; by restraining the rules for these modalities, one achieves the desired goal. Of course one has to make sure that the resulting system is also \emph{complete}, i.e., that \emph{all} functions of the given complexity can be programmed in it. This methodology has been successfully followed to characterize complexity classes like deterministic polynomial time \citep{GirardScedrovScott92,Girard:LLL,AspertiRoversi02,Lafont:SLL}, elementary time \citep{Girard:LLL,DanosJoinet:ELL}, deterministic logarithmic space \citep{Schoepp07}, and, very recently, polynomial space \citep{GaboardiMarionRonchi:PSpace}.}

\paragraph*{Stratification.} In this work we focus on elementary linear logic (\ELL) and light linear logic (\LLL) systems, corresponding to elementary time and deterministic polynomial time, respectively \cite{Girard:LLL}.

The complexity bound on the cut-elimination procedure of these systems relies on a principle called \emph{stratification}, which is also at the base of other approaches to implicit computational complexity, both related to logic and not.

Stratification can be interpreted in at least three informal ways. The first, which is where \cite{Girard:LLL} originally drew inspiration from, comes from a sharp analysis of Russel's paradox in naive set theory \citep{Fitch,CurryFeys:CombLogic}, and was first considered by \cite{Leivant94a}. Unrestricted comprehension can be obtained as a theorem in first order classical logic plus the following two rules:
\TwoRules
{\infer{\ts\Gamma,t\in\{x~|~A\}}{\ts\Gamma,A[t/x]}} {\infer{\ts\Gamma,t\not\in\{x~|~A\}}{\ts\Gamma,\lnot A[t/x]}}
where $\{x~|~A\}$ is the standard set-builder notation for the set containing all and only the elements satisfying the formula $A$. Russel's antinomy is obtained by considering the term $r=\{x~|~x\not\in x\}$, from which we build the formula $R=r\in r$. One can see that $R$ is a fixpoint of negation, i.e., $R$ is provably equivalent to $\lnot R$. In fact, one can obtain $\ts\Gamma,R$ from $\ts\Gamma,\lnot R$ by applying the rule above on the left, and $\ts\Gamma,\lnot R$ from $\ts\Gamma,R$ by applying the rule above on the right. The empty sequent, i.e., a contradiction, can then be derived as follows:
\begin{center}
	\mbox{
		\infer{\ts}
		{
			\infer{\ts\lnot R}
			{
				\infer{\ts\lnot R,\lnot R}{\infer{\ts\lnot R,R}{}}
			}
			&
			\infer{\ts R}
			{
				\infer{\ts R,R}{\infer{\ts\lnot R,R}{}}
			}
		}
	}
\end{center}
Remark that contraction is necessary: in multiplicative linear logic, where contraction is forbidden, the empty sequent cannot be derived even in presence of the self-contradicting formula $R$ (this was first observed by \cite{Grishin}).

Another setting in which stratification can be applied is the \lac, where Russel's paradox corresponds to the diverging term $\Omega$. The fundamental construct behind this term is self-application, which, from the logical point of view, also needs contraction.

A third intuition comes from recursion theory, where more and more complex functions can be obtained by diagonalization. For instance, if $P_m(n)$ is a sequence of polynomial functions of degree $m$ in $n$ (for example, $P_m(n)=n^m$), the function $P_n(n)$ is super-exponential, i.e., elementary; if $\theta_m(n)$ is a sequence of elementary functions in $n$ whose complexity rises with $m$ (for example, $\theta_m(n)=2_m^n$, i.e., a tower of exponentials of height $m$ in $n$), then $\theta_n(n)$ is hyper-exponential, i.e., non-elementary.

In all of these incarnations, stratification can be seen as a way of forbidding the identification of two variables, or the contraction of two formulas, because they belong to two morally different ``levels'': the occurrence of $R$ coming from the axiom and that coming from the application of the $\in$-rule in the derivation of Russel's paradox; the occurrence of $x$ in function position and that in argument position in the self application $\lambda x.xx$; the index of the sequence and the argument of the members of the sequence in the diagonalization examples.

Note that stratification is reminiscent of the notion of \emph{ramification}, or its variants like \textit{safe recursion}, used for restricting primitive recursion in implicit computational complexity~\citep{BellantoniCook92,LeivantMarion93,Leivant94}. The relation between safe recursion and light linear logic was investigated in~\cite{MurawskiOng04}, while a study on diagonalization and complexity was recently carried out by \cite{Marion:Diag}.

\paragraph*{\Pn s, boxes, and stratification.} The bound on the cut-elimination procedure for \ELL\ and \LLL\ is proved using \emph{\pn s}, a graphical representation of proofs~\citep{Girard:ProofNets}. These are a crucial tool for applying linear logic to implicit computational complexity: they allow a fine-grained analysis of cut-elimination, the definition of adequate measures and invariants, and the introduction of adapted reduction strategies. In particular, the fundamental stratification property of \ELL\ and \LLL\ is defined and enforced through \emph{boxes}, a construct in the syntax of \pn s corresponding to the rules for exponential modalities. Boxes have been around since the introduction of \pn s \citep{Girard:LL} and can be understood intuitively in two ways:
\begin{enumerate}
	\item[(i)] \emph{logically}: they correspond to sequentiality information;
	\item[(ii)] \emph{operationally}: they mark subgraphs (i.e., subproofs) that can be duplicated.
\end{enumerate}

Boxes can be nested; as a consequence, a node in a \pn\ (corresponding to a logical rule) may be assigned an \emph{exponential depth}, which is the number of nested boxes containing that node. Stratification is achieved precisely on the base of the exponential depth: in full linear logic, two occurrences of the same formula introduced at different exponential depths may eventually be contracted; in \ELL\ and \LLL, they cannot. From the operational point of view, boxes therefore assume a twofold role in \ELL\ and \LLL: they serve for the purpose (ii) explained above, and they enforce stratification.

\paragraph*{A new stratification.} The main contribution of this work is the investigation of an alternative way to achieve stratification, which is orthogonal to boxes. It is a direct application of the intuitions concerning stratification given above: occurrences of formulas in a \pn\ are ``tested'' by assigning to them an \emph{index}, which must satisfy certain constraints; in particular, if two occurrences of the same formula are contracted, then they must have the same index. If the \pn\ ``passes the test'', i.e., if there is a way of assigning indexes to its formulas in a way which is compatible with the constraints, then the \pn\ is accepted.

The assignment of indexes naturally determines the stratification of a proof net into \textit{levels}, which need not match exponential depths. We thus define a system called \emph{linear logic by levels} (\LLlev), prove that it admits an elementary bound on cut elimination, and that it is complete for elementary time functions. It actually turns out that \ELL\ corresponds to the subsystem of \LLlev\ in which levels and depths coincide, so finally Girard's approach to stratification can be seen as a special case of our own.

The idea of using indexes in linear logic proofs can already be found in the work on \emph{2-sequent calculi} by \cite{Masini:TwoSeq} and \cite{GuerriniMartiniMasini:TwoSeq}. In the latter paper the authors define 2-sequent calculi systems corresponding to \ELL\ and \LLL. However, our goal here is different because we are not primarily interested in reformulating \ELL\ and \LLL\ but rather in generalizing these systems and proving properties directly for such generalizations.

As said above, the main novelty of \LLlev\ is that it shows how stratification and exponential depths must not necessarily be related. This is, in our opinion, an important contribution to the understanding of the principles underlying light logics. It may also be a starting point for finding new kinds of denotational semantics for bounded time computation, extending the ideas of \cite{Baillot04a} and \cite{LaurentTortora:OCliques}.

\paragraph*{Removing useless boxes.} In \LLL, along the exponential modalities of linear logic, an additional exponential modality, the \emph{paragraph}~$\S$, must be added in order to reach the desired expressive power, i.e., programming all polytime functions. Since stratification is linked to exponential depth, the paragraph modality too is handled in \pn s by means of boxes; however, $\S$-boxes cannot be duplicated, so they lose their original function (ii), and their existence is only justified by stratification.

By imposing on our \LLlev\ the same kind of constraints that define \LLL\ from \ELL, we obtain \emph{light linear logic by levels} (\LLLlev), which, as expected, characterizes deterministic polynomial time. This system offers an additional advantage with respect to \LLL: since our stratification is orthogonal to boxes, and since \mbox{$\parg$-boxes} exist only to enforce stratification, these are no longer needed in \LLLlev.

\paragraph*{Improving type systems.} In several cases, the characterization of complexity classes with subsystems of linear logic has allowed, in a second step, to define type systems for the \lac\ statically ensuring complexity properties~\citep{BaillotTerui:DLAL,GaboardiRonchi07}: if a \lat, expecting for instance a binary list argument, is well typed, then it admits a complexity bound w.r.t.\ the size of the input. Such results naturally call for type inference procedures~\citep{coppola06tocl,AtassiBaillotTerui07}, which can be seen as tests for sufficient conditions for a program to admit a complexity bound.

From this point of view, the presence of \mbox{$\S$-boxes} in \LLL\ is a heavy drawback: in fact, a large part of the work needed to perform type inference in \LLL, or subsystems like \DLAL~\citep{AtassiBaillotTerui07}, comes from the problem of placing correctly \mbox{$\S$-boxes}, in particular in such a way that they are compatible with other rules, or with $\lambda$ bindings in the \lac\ (remember that boxes also carry sequentialization information, cf.\ point (i) above). A system like \LLLlev\ clearly offers the possibility of overcoming these problems: the absence of \mbox{$\S$-boxes} may yield major simplifications in the development of type systems for polynomial time.

A further contribution of this paper is making a first step in that direction: exploiting the lack of sequentiality constraints on the paragraph modality, we devise a variant of \LLLlev\ in which the paragraph modality is hidden in atomic formulas; as a consequence, the paragraph modality completely disappears from this system, and there is no need for a rule handling it. This may turn out to be  extremely helpful for designing a type system out of our work.

\paragraph*{Plan of the paper.} \refsect{MELL} contains a sort of mini-crash-course on linear logic and its light subsystems \ELL\ and \LLL. Apart from introducing the material necessary to our work, this (quite lengthy) section should make the paper as self-contained as possible, and hopefully accessible to the reader previously unfamiliar with these topics. The systems \LLlev\ and \LLLlev\ are introduced in \refsect{LLlev}, and their relationship with \ELL\ and \LLL\ is spelled out. \refsect{Complexity} is the technical core of the paper: it contains the proof of the complexity bounds for \LLlev\ (\refth{ElemBound}) and \LLLlev\ (\refth{PolyBound}), from which the characterization result follows (\refth{Characterization}). \refsect{LLLlevz} introduces the variant of \LLLlev\ without paragraph modality; the main result of this section is \refth{Characterizationz}. In \refsect{Conclusions} we conclude the paper with a discussion about open questions and future work.

\paragraph*{Acknowledgments.} The authors would like to thank Daniel de Carvalho for his useful comments and suggestions on the subject of this paper. This work was partially supported by project NOCoST (ANR, JC05\_43380).

\section{Multiplicative Exponential Linear Logic}
\label{sect:MELL}

\subsection{Formulas}
\label{sect:Formulas}
The formulas of second order unit-free multiplicative exponential linear logic (\MELL) are generated by the following grammar, where $X,X^\perp$ range over a denumerable set of propositional variables:
$$A,B::=X~|~X^\perp~|~A\ltens B~|~A\lpar B~|~\oc A~|~\wn A~|~\exists X.A~|~\forall X.A~|~\S A.$$
Linear negation is defined through De Morgan laws:
\begin{displaymath}
	\begin{array}{rclcrcl}
		(X)^\perp & = & X^\perp && (X^\perp)^\perp & = & X \\
		(A\ltens B)^\perp & = & B^\perp\lpar A^\perp && (A\lpar B)^\perp & = & B^\perp\ltens A^\perp \\
		(\oc A)^\perp & = & \wn A^\perp && (\wn A)^\perp & = & \oc A^\perp\\
		(\exists X.A)^\perp & = & \forall X.A^\perp && (\forall X.A)^\perp & = & \exists X.A^\perp \\
		&&& (\parg A)^\perp\ =\ \parg A^\perp &&&
	\end{array}
\end{displaymath}
Two connectives exchanged by negation are said to be \emph{dual}. Note that the self-dual paragraph modality is not present in the standard definition of \MELL\ \citep{Girard:LL}; we include it here for convenience. Also observe that full linear logic has a further pair of dual binary connectives, called \emph{additive} (denoted by $\&$ and $\oplus$), which we shall briefly discuss in \refsect{Additives}. They are not strictly needed for our purposes, hence we restrict to \MELL\ in the paper.

Linear implication is defined as $A\llinimp B=A^\perp\lpar B$. Multisets of formulas will be ranged over by $\Gamma,\Delta,\ldots$

For technical reasons, it is also useful to consider \emph{\struct s}, which will be denoted by $\flat A$, where $A$ is a formula.

\subsection{Proofs}
\label{sect:Proofs}
\begin{table}[t]
	\TwoRules
	{\infer[\etq{Axiom}]{\ts A^\perp,A}{}}
	{\infer[\etq{Cut}]{\ts\Gamma,\Delta}{\ts\Gamma,A & \ts\Delta,A^\perp}}
	\smallskip
	\TwoRules
	{\infer[\etq{Tensor}]{\ts\Gamma,\Delta,A\ltens B}{\ts\Gamma,A & \ts\Delta,B}}
	{\infer[\etq{Par}]{\ts\Gamma,A\lpar B}{\ts\Gamma,A,B}}
	\smallskip
	\TwoRules
	{\infer[\etq{For all ($X$ not free in $\Gamma$)}]{\ts\Gamma,\forall X.A}{\ts\Gamma,A}}
	{\infer[\etq{Exists}]{\ts\Gamma,\exists X.A}{\ts\Gamma,A[B/X]}}
	\smallskip
	\TwoRules
	{\infer[\etq{Promotion}]{\ts\wn\Gamma,\oc A}{\ts\wn\Gamma,A}}
	{\infer[\etq{Dereliction}]{\ts\Gamma,\wn A}{\ts\Gamma,A}}
	\smallskip
	\TwoRules
	{\infer[\etq{Weakening}]{\ts\Gamma,\wn A}{\ts\Gamma}}
	{\infer[\etq{Contraction}]{\ts\Gamma,\wn A}{\ts\Gamma,\wn A,\wn A}}
	\smallskip
	\OneRule
	{\infer[\etq{Paragraph}]{\ts\Gamma,\parg A}{\ts\Gamma,A}}
	\smallskip
	\TwoRules
	{\infer[\etq{Daimon}]{\ts}{}}
	{\infer[\etq{Mix}]{\ts\Gamma,\Delta}{\ts\Gamma & \ts\Delta}}
	\caption{The rules for \MELL\ sequent calculus.}
	\label{tab:MELLSeq}
\end{table}
\paragraph*{Sequent calculus and cut-elimination.} The proof theory of \MELL\ can be formulated using the sequent calculus of \reftab{MELLSeq}. This calculus, which can be shown to enjoy cut-elimination, differs from the one originally given by \cite{Girard:LL} because of the addition of the last three rules. All of them are added for convenience. The paragraph rule actually makes this modality trivial, as expressed by the following:
\begin{prop}
	\label{prop:PargIso}
	For any $A$, $\S A$ is provably isomorphic to $A$ in \MELL.
\end{prop}
\begin{pf}
	It is not hard to see that there are two derivations $D_1,D_2$ of $\ts\S A^\perp,A$ and $\ts A^\perp,\S A$, from which one can obtain two derivations of $\ts\S A\llinimp A$ and $\ts A\llinimp\S A$, respectively. Moreover, the derivations obtained by cutting $D_1$ with $D_2$ in the two possible ways both reduce to the identity (i.e., an axiom modulo $\eta$-expansion) after cut-elimination.\qed
\end{pf}
Nevertheless, we shall consider subsystems of \MELL\ in which the paragraph modality is not trivial, and this is why we find it convenient to include it right from the start. The mix rule, and its nullary version (here called the daimon rule), are discussed more thoroughly at the end of this section. Basically, their presence simplifies the presentation of \pn s.

This last point is very important to us. In fact, the backbone of our work is a detailed analysis, in terms of computational complexity, of the cut-elimination procedure of \MELL. In sequent calculus, this is composed of rules which are suitable reformulations of those originally given by \cite{Gentzen:CutElim} to prove his \emph{Hauptsatz} for classical logic (the calculus $\mathbf{LK}$). As a consequence, most of them are commutations, i.e., rules permuting a cut with another inference rule; only a few of them act on derivations in a non-trivial way. This is why we consider \pn s, an alternative presentation of the proof theory of \MELL\ offering, among other things, the advantage of formulating cut-elimination without commutations: only the ``interesting'' rules are left.

\paragraph*{\Pn s.} The \pn\ formalism was introduced by \cite{Girard:LL,Girard:ProofNets}, and subsequently reformulated by other authors using slightly different syntactical definitions. In this paper, we use a combination of the presentations given by \cite{DanosRegnier:PNHilb} and \cite{Tortora:AdditivesAndNormalizationI}, with a slight change in the terminology: the term ``proof structure'', introduced by~\cite{Girard:LL} and traditionally used in the literature, is here dismissed in favor of the term \emph{\ps}. On the contrary, the term \pn, i.e., a \ps\ satisfying certain structural conditions (the correctness criterion), retains its usual meaning.

\begin{figure}[t]
	\begin{center}\scalebox{\scalefact}{\input{links.tex}}\end{center}
	\caption{Links.}
	\label{fig:Links}
\end{figure}
\begin{figure}[t]
	\begin{center}\scalebox{\scalefact}{\input{boxes.tex}}\end{center}
	\caption{A box.}
	\label{fig:Boxes}
\end{figure}
In the following definition, and throughout the rest of the paper, unless explicitly stated we shall make no distinction between the concepts of \emph{formula} and \emph{occurrence of formula}. The same will be done for what we call \emph{links} and their occurrences.
\begin{defn}[\Ps]
	\label{def:ProofStructure}
	A \emph{\pps} is a pair $(\mathcal G,\mathsf B)$, where $\mathcal G$ is a finite graph-like object whose nodes are occurrences of what we call \emph{links}, and whose edges are directed and labelled by formulas or \struct s of \MELL; and $\mathsf B$ is a set of subgraphs of $\mathcal G$ called \emph{boxes}.
	\begin{itemize}
		\item Links (\reffig{Links}) are labelled by connectives of \MELL, or by one of the labels \axlink, \cutlink, $\flat$, \paxlink. Two links labelled by dual connectives are said to be \emph{dual}. Each link has an arity and co-arity, which are resp.\ the number of its incoming and outgoing edges. The arity and co-arity is fixed for all links except \whynot\ links, which have co-arity $1$ and arbitrary arity. A nullary \whynot\ link is also referred to as a \weak\ link. $\mathsf{Par}$ and \univ\ links are called \emph{jumping} links.
		
		\item The incoming edges of a link (and the formulas that label them) are referred to as its \emph{premises}, and are assumed to be ordered, with the exception of \cut\ and \whynot\ links; the outgoing edges of a link (and the formulas that label them) are referred to as its \emph{conclusions}.
		
		\item Premises and conclusions of links must respect a precise labeling (which depends on the link itself), given in \reffig{Links}. In particular:
		\begin{itemize}
			\item edges labelled by \struct s can only be premises of \paxl\ and \whynot\ links;
			\item in a \univ\ link $l$, the variable $Z$ in its premise $A[Z/X]$ is called the \emph{eigenvariable} of $l$. Each \univ\ link is assumed to have a different eigenvariable.
			\item in an \exist\ link $l$, the formula $B$ in its premise $A[B/X]$ is said to be \emph{associated} with $l$.
		\end{itemize}
		
		\item Each edge must be the conclusion of exactly one link, and the premise of at most one link. The edges that are not premises of any link (and the formulas that label them) are deemed \emph{conclusions} of the \pps. (Note that the presence of these ``pending'' edges, together with the fact that some premises are ordered, is why \pps s are not exactly graphs).
		
		\item A box is depicted as in \reffig{Boxes}, in which $\pi$ is a \pps, said to be \emph{contained} in the box. The links that are explicitly represented in \reffig{Boxes} (i.e., the \paxl\ links and the \ofcourse\ link) form the \emph{border} of the box. The unique \ofcourse\ link in the border is called the \emph{\pal} of the box, while the \paxl\ links are called \emph{\pax s}. We have the following conditions concerning boxes:
		\begin{enumerate}
			\item[a.] each \ofcourse\ link is the \pal\ of exactly one box;
			\item[b.] each \paxl\ link is in the border of exactly one box; 
			\item[c.] any two distinct boxes are either disjoint or included in one another.
		\end{enumerate}
	\end{itemize}
	
	A \emph{\ps} is a \pps\ such that in its conclusions there is no \struct, nor any formula containing an eigenvariable.
\end{defn}
\begin{defn}[Depth, size]
	\label{def:DepthSize}
	Let $\sigma$ be a \pps.
	\begin{itemize}
		\item A link (or edge) of $\sigma$ is said to have \emph{depth} $d$ if it is contained in $d$ (necessarily nested) boxes. The depth of a box of $\sigma$ is the depth of the links forming its border. The depth of a link $l$, edge $e$, or box $\cB$ are denoted resp.\ by $\Depth l$, $\Depth e$ and $\Depth\cB$. The depth of $\sigma$, denoted by $\Depth\sigma$, is the maximum depth of its links.
		\item The \emph{size} of $\sigma$, denoted by $\Size\sigma$, is the number of links contained in $\sigma$, excluding \pax s.
	\end{itemize}
\end{defn}
\begin{defn}[Switching]
	\label{def:Switching}
	Let $\sigma$ be a \pps. For each jumping link $l$ of $\sigma$, we define the set of \emph{jumps} of $l$, denoted by $J(l)$, as follows:
	\begin{description}
		\item[\parl:] $J(l)$ is the set containing the links whose conclusions are the premises of~$l$.
		\item[\univ:] if $Z$ is the eigenvariable of $l$, $J(l)$ is the set containing:
		\begin{itemize}
			\item the link whose conclusion is the premise of $l$;
			\item any link whose conclusion is labelled by a formula containing $Z$;
			\item any \exist\ link whose associated formula contains $Z$.
		\end{itemize}
	\end{description}
	A \emph{switching} of $\sigma$ is an undirected graph built as follows:
	\begin{itemize}
		\item the conclusions of $\sigma$ are erased, and its edges considered as undirected;
		\item for each jumping link $l$, the premises of $l$ (if any) are erased, exactly one node $m\in J(l)$ is chosen and a new edge between $m$ and $l$ is added.
		\item the boxes at depth zero of $\sigma$ are collapsed into single nodes, i.e., if $\cB$ is a box at depth zero of $\sigma$, it is erased together with all the edges connecting its links to the rest of the graph, and replaced with a new node $l$; then, for any link $m$ of depth zero which was connected to a link of $\cB$, a new edge between $m$ and $l$ is added.
	\end{itemize}
\end{defn}
\begin{defn}[\Pn]
	\label{def:ProofNet}
	A \pps\ $(\mathcal G,\mathsf B)$ is \emph{correct} iff:
	\begin{itemize}
		\item all of its switchings are acyclic;
		\item for all $\cB\in\mathsf B$, the \pps\ contained in $\cB$ is correct.
	\end{itemize}
	A \emph{\pn} is a correct \ps.
\end{defn}

\paragraph*{Sequent calculus and \pn s.} The relationship between sequent calculus and \pn s is clarified by the notion of \emph{sequentializable} \ps, whose definition mimics the rules of sequent calculus:
\begin{figure}[t]
	\begin{center}\scalebox{1}{\input{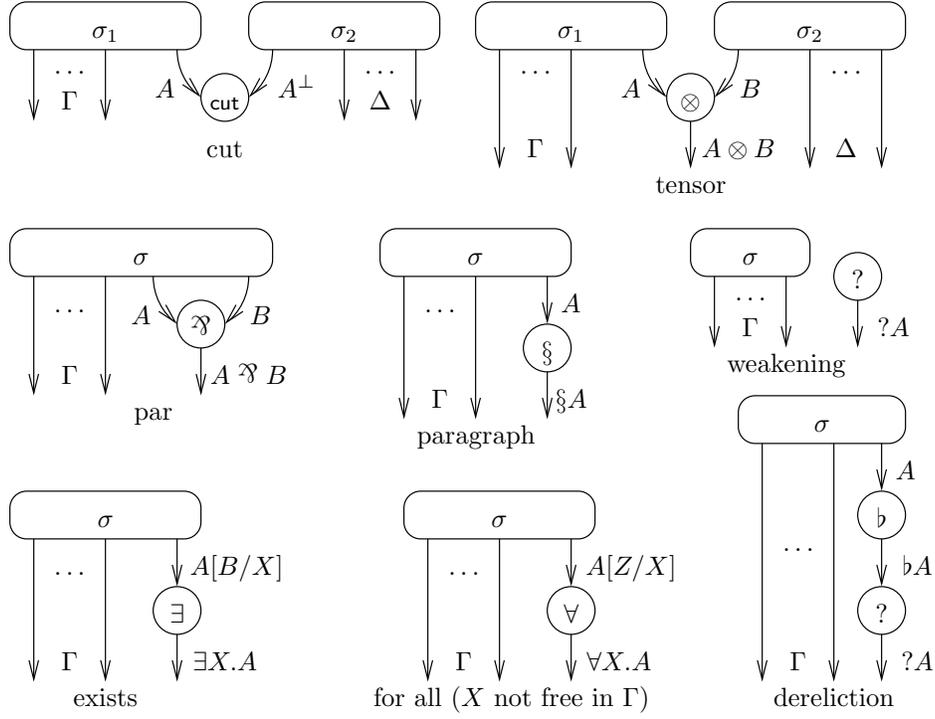}}\end{center}
	\caption{Rules for building sequentializable \ps s.}
	\label{fig:SPS}
\end{figure}
\begin{defn}[\Sps]
	\label{def:SeqProofStructure}
	We define the set of \emph{\sps s} inductively: the empty \ps\ and the \ps\ consisting of a single \axiom\ link are sequentializable (daimon and axiom); the juxtaposition of two sequentializable \ps s is sequentializable (mix); if $\sigma$, $\sigma_1$, $\sigma_2$ are \sps s of suitable conclusions, the \ps s of \reffig{SPS} are sequentializable; if
	\begin{center}\scalebox{\scalefact}{\input{PromotionSPS1.tex}}\end{center}
	is a \sps, then the \ps
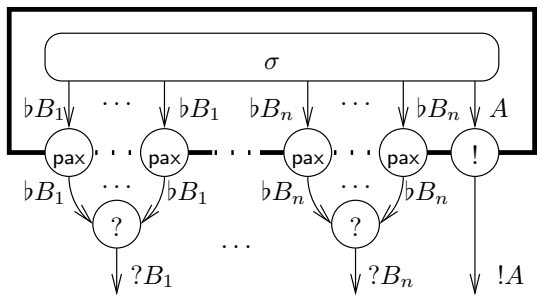
	\begin{center}\scalebox{\scalefact}{\input{PromotionSPS2.tex}}\end{center}
	is sequentializable (promotion); if
	\begin{center}\scalebox{\scalefact}{\input{ContractionSPS1.tex}}\end{center}
	is a \sps, then the \ps
	\begin{center}\scalebox{\scalefact}{\input{ContractionSPS2.tex}}\end{center}
	is sequentializable (contraction).
\end{defn}

\begin{prop}[\cite{Girard:ProofNets}]
	\label{prop:Sequentialization}
	A \ps\ is sequentializable iff it is a \pn.
\end{prop}
The above result, combined with \refdef{SeqProofStructure}, gives a simple intuition for looking at \pn s: they can be seen as a sort of ``graphical sequent calculus''.

\paragraph*{Cut-elimination.}
\begin{figure}[p]
	\begin{center}\scalebox{\scalefact}{\input{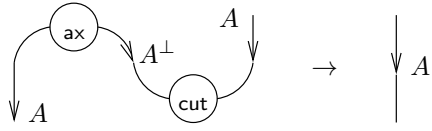}}\end{center}
	\caption{Axiom step.}
	\label{fig:AxStep}
\end{figure}
\begin{figure}[p]
	\begin{center}\scalebox{\scalefact}{\input{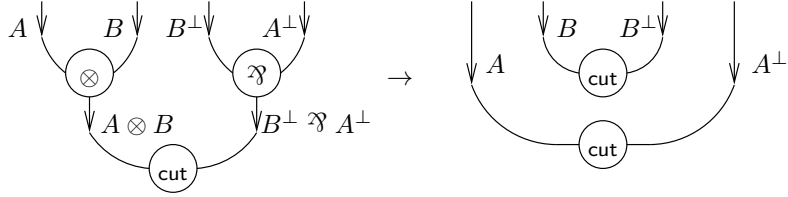}}\end{center}
	\caption{Multiplicative step.}
	\label{fig:MultStep}
\end{figure}
\begin{figure}[pt]
	\begin{center}\scalebox{\scalefact}{\input{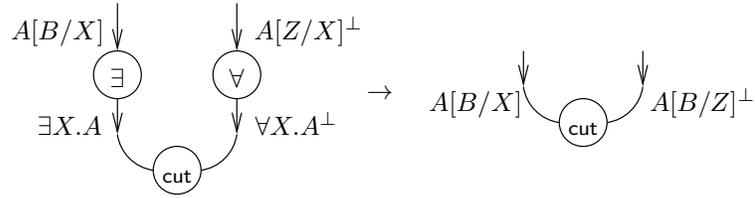}}\end{center}
	\caption{Quantifier step; the substitution is performed on the whole net.}
	\label{fig:QuantStep}
\end{figure}
\begin{figure}[pt]
	\begin{center}\scalebox{0.7}{\input{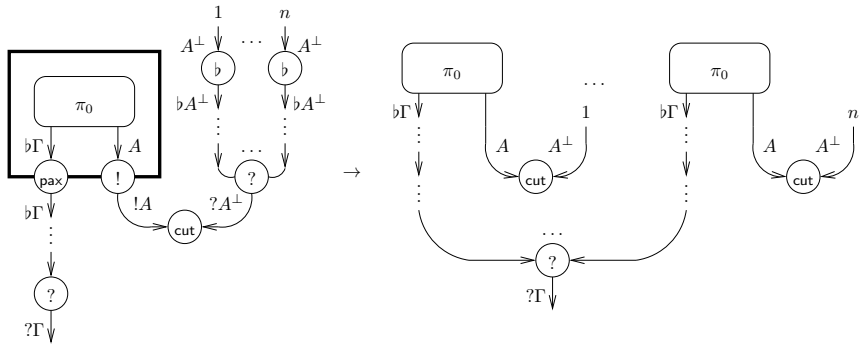}}\end{center}
	\caption{Exponential step; $\flat\Gamma$ is a multiset of \struct s, so one \paxl\ link, \whynot\ link, or wire in the picture may in some case stand for several (including zero) \paxl\ links, \whynot\ links, or wires.}
	\label{fig:ExpStep}
\end{figure}
\begin{figure}[pt]
	\begin{center}\scalebox{\scalefact}{\input{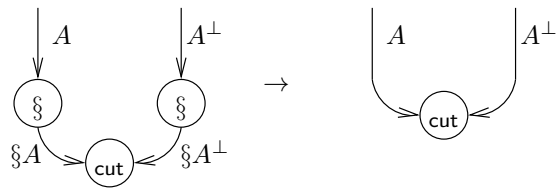}}\end{center}
	\caption{Paragraph step.}
	\label{fig:PargNoBoxStep}
\end{figure}
As anticipated above, formulating the cut-elimination procedure in \pn s is quite simple: there are only five rules (or \emph{steps}, as we shall more often call them), taking the form of the graph-rewriting rules given in Figures~\ref{fig:AxStep} through \ref{fig:PargNoBoxStep}. When a \ps\ $\pi$ is transformed into $\pi'$ by the application of one cut-elimination step, we write $\pi\onered\pi'$, and we say that $\pi$ \emph{reduces} to $\pi'$. Of course, in that case, if $\pi$ is a \pn, then $\pi'$ is also a \pn, i.e., cut-elimination preserves correctness.

The following notions, taken from \cite{Tortora:AdditivesAndNormalizationI}, are needed to analyze the dynamics of \pn s under cut-elimination, and will prove to be quite useful in the sequel:
\begin{defn}[Lift, residue]
	Whenever $\pi\onered\pi'$, by simple inspection of the cut-elimination rules it is clear that any link $l'$ of $\pi'$ different from a \cut\ comes from a unique (``the same'') link $l$ of $\pi$; we say that $l$ is the \emph{lift} of $l'$, and that $l'$ is a \emph{residue} of $l$. We define the lift and residues of a box in the same way.
\end{defn}

\paragraph*{Untyped \pn s.} We shall also use an untyped version of \pn s:
\begin{defn}[Untyped \pn]
	\label{def:UProofNet}
	An untyped \pps\ is a directed graph with boxes built using the links of \reffig{Links} as in \refdef{ProofStructure}, but without any labels on edges, or any constraint induced by such labels. An untyped \ps\ is an untyped \pps\ such that:
	\begin{itemize}
		\item the conclusion of a \flatl\ link must be the premise of a \paxl\ or \whynot\ link;
		\item the premise of a \paxl\ link must be the conclusion of a \flatl\ or \paxl\ link, and the conclusion of \paxl\ link must be the premise of a \paxl\ or \whynot\ link;
		\item the premises of a \whynot\ link must be conclusions of \flatl\ or \pax\ links.
	\end{itemize}
	The notion of switching can be applied to untyped \pps s with virtually no change (\univ\ links are no more jumping links), and hence the notion of correctness. We then define an untyped \pn\ as a correct untyped \ps.
\end{defn}

Cut-elimination can be defined also for untyped \ps s. In fact, of all cut-elimination steps, only the quantifier step (\reffig{QuantStep}) actually uses formulas; however, even in this case the modifications made to the underlying untyped \ps\ do not depend on formulas. Hence, in the untyped case, the quantifier step and the paragraph step (\reffig{PargNoBoxStep}) behave identically. Obviously, in the untyped case there may be ``clashes'', i.e., \cut\ links connecting the conclusions of two non-dual links. In that case, the \cut\ link is said to be \emph{irreducible}; otherwise, we call it \emph{reducible}. Hence, untyped \pn s may admit normal forms which are not cut-free.

\paragraph*{Remarks on mix and daimon.} We mentioned above that admitting the mix and daimon rules makes the definition of \pn s simpler. In fact, at present, all known solutions excluding them are quite cumbersome and bring up issues which are morally unproblematic but technically disturbing \cite{Tortora:AdditivesAndNormalizationI}.

The status of the mix rule in the proof theory of linear logic is somewhat controversial~\citep{Girard:LePointAveugle}. Its computational meaning is not clear, and no complexity-related subsystem of linear logic makes use of it. Its presence is harmless though: as a matter fact, while we shall explicitly rely on the acyclicity condition of \refdef{ProofNet} in one crucial occasion (\reflemma{POrd}), the soundness of our systems (Theorems~\ref{th:ElemBound}~and~\ref{th:PolyBound}) holds without requesting any further condition on switchings which would exclude daimon or mix. Nevertheless, the completeness results (\refsect{Characterization}) hold for much smaller subsystems, using none of the debated rules (see \refsect{IntuitionLambda} below). For this reason, the reader who is puzzled by daimon and mix (in particular the former, which makes the empty sequent provable in \MELL, and with it all formulas of the form $\wn A$) may simply forget about their existence.

\subsection{Computational interpretation}
\label{sect:IntuitionLambda}
The most direct computational interpretation of \MELL\ can be given by considering its intuitionistic subsystem. The intuitionistic (or, more precisely, minimal) sequent calculus of \MELL\ is obtained from that of \reftab{MELLSeq} in the same way one obtains $\mathbf{LJ}$ from $\mathbf{LK}$ \citep{Gentzen:CutElim}. The interest of the intuitionistic sequent calculus for \MELL\ is that its derivations can be decorated with \lat s in such a way that cut-elimination in proofs is consistent with \mbox{$\beta$-reduction} in the \lac.

\begin{table}[t]
	\TwoRules
	{\infer[\etq{Axiom}]{x:A\ts x:A}{}}
	{\infer[\etq{Cut}]{\Gamma\ts u[t/x]:B}{\Gamma\ts t:A & \Delta,x:A\ts u:B}}
	\smallskip
	\TwoRules
	{\infer[\etq{R$\llinimp$}]{\Gamma\ts \lambda x.u:A\llinimp B}{\Gamma,x:A\ts u:B}}
	{\infer[\etq{L$\llinimp$}]{\Gamma,\Delta,z:A\llinimp B\ts v[zt/y]}{\Gamma\ts t:A & \Delta,y:B\ts v:C}}
	\smallskip
	\TwoRules
	{\infer[\etq{L$\forall$}]{\Gamma,x:\exists X.A\ts u:B}{\Gamma,x:A[B/X]\ts u:B}}
	{\infer[\etq{R$\forall$ ($X$ not free in $\Gamma$)}]{\Gamma\ts t:\forall X.A}{\Gamma\ts t:A}}
	\smallskip
	\TwoRules
	{\infer[\etq{D}]{\Gamma,x:\oc A\ts u:B}{\Gamma,x:A\ts u:B}}
	{\infer[\etq{P}]{\oc\Gamma\ts t:\oc A}{\oc\Gamma\ts t:A}}
	\smallskip
	\TwoRules
	{\infer[\etq{W}]{\Gamma,x:\oc A\ts u:B}{\Gamma\ts u:B}}
	{\infer[\etq{C ($z$ fresh)}]{\Gamma,z:\oc A\ts u[z/x,z/y]:B}{\Gamma,x:\oc A,y:\oc A\ts u:B}}
	\smallskip
	\TwoRules
	{\infer[\etq{L$\parg$}]{\Gamma,x:\parg A\ts u:B}{\Gamma,x:A\ts u:B}}
	{\infer[\etq{R$\parg$}]{\Gamma\ts t:\parg A}{\Gamma\ts t:A}}
	\caption{The rules for \MELL\ intuitionistic sequent calculus, and their attached \lat s.}
	\label{tab:IntLambdaSeq}
\end{table}
The calculus is given in \reftab{IntLambdaSeq}, directly with the decorations. Note that, as expected, the constraint of having exactly one formula to the right of sequents suggests to treat linear implication as a primitive connective, and to eliminate the par connective. For the same reason, the daimon and mix rules are excluded.

By translating $A\llinimp B$ as $A^\perp\lpar B$, and by converting an intuitionistic sequent $\Gamma\ts A$ into $\ts\Gamma^\perp,A$, one can define \emph{intuitionistic \pn s} as \ps s which can be built mimicking the rules of \reftab{IntLambdaSeq}, in the spirit of \refdef{SeqProofStructure}. Intuitionistic \pn s are of course \pn s, but the decoration of \reftab{IntLambdaSeq} attaches a \lat\ to them. As anticipated above, this turns into a concrete computational semantics, thanks to the following:
\begin{prop}
	\label{prop:ComputSem}
	Let $\pi$ be an intuitionistic \pn, and let $\pi\onered\pi'$. Then:
	\begin{enumerate}
		\item $\pi'$ is intuitionistic;
		\item if $t,t'$ are the \lat s attached to $\pi,\pi'$, respectively, then $t\red_\beta t'$.
	\end{enumerate}
\end{prop}

\refprop{ComputSem} is a useful guideline for programming with \MELL\ \pn s: if one sticks to the intuitionistic subsystem, it is possible to use the \lac\ as a target language into which \pn s can be ``compiled''. All complexity-related subsystems of \MELL\ exploit this; as a matter of fact, the completeness with respect to the complexity classes they characterize is always proved within their intuitionistic subsystem. This will be the case for our systems too.

\subsection{Elementary and light linear logic}
\label{sect:ELLandLLL}
The logical systems which are the main objects of this paper are extensions of the multiplicative fragments of elementary linear logic (\ELL) and light linear logic (\LLL), both introduced by \cite{Girard:LLL}. These two systems characterize, in a sense which will be made precise at the end of the section, the complexity classes \FE\ and \FP, respectively: the former is the class of functions computable by a Turing machine whose runtime is bounded by a tower of exponentials of fixed height (also known as \emph{elementary functions}); the latter is the class of functions computable in polynomial time by a deterministic Turing machine. In this section, we briefly recall the definition of these two systems.

\paragraph*{The stratification condition.} The multiplicative fragment of \ELL\ can be defined in our \pn\ syntax by using the notion of \emph{exponential branch}, as in \cite{DanosJoinet:ELL}:
\begin{defn}[Exponential branch]
	\label{def:ExpBranch}
	\sloppy{Let $\sigma$ be a (typed or untyped) \MELL\ \ps, and let $b$ be a \flatl\ link of $\sigma$.  The \emph{exponential branch} of $b$ is the directed path starting from the conclusion of $b$, crossing a number (maybe null) of auxiliary ports and ending in the premise of a \whynot\ link (which must exist by \refdef{ProofStructure}, or \refdef{UProofNet} in the untyped case).}
\end{defn}
\begin{defn}[Multiplicative elementary linear logic]
	\label{def:ELL}
	\sloppy{Multiplicative elementary linear logic (\MultELL) is the subsystem of \MELL\ composed of all \pn s satisfying the following condition:}
	\begin{description}
		\item[Depth-stratification:] Each exponential branch of $\pi$ crosses exactly one auxiliary port.
	\end{description}
\end{defn}
Note once again that the paragraph modality is absent in original definition of \MultELL, but including it is harmless (\refprop{PargIso} still holds).

Of course the depth-stratification condition is preserved by cut-elimination: if $\pi$ is in \MultELL, and $\pi\onered\pi'$, then $\pi'$ is also in \MultELL. As suggested by its name, the fundamental purpose of this condition is to assure a \emph{stratification property}, which can be formally stated as follows: whenever $\pi\onered\pi'$, if $l$ is a link of $\pi$ different from a \cut\ and $l'$ is a residue of $l$ in $\pi'$, we have $\Depth{l'}=\Depth{l}$. By contrast, in a generic \MELL\ \pn\ a residue of a link $l$ may also have depth smaller (by one) or greater (by any number) than $l$ itself. In other words, depths can ``communicate'' in \MELL, but are ``separated worlds'' in \MultELL.

\paragraph*{Round-by-round cut-elimination.} \sloppy{The essential property of a \MultELL\ \pn\ $\pi$ is that its cuts can be eliminated so that the size of all \pn s obtained during cut-elimination is bounded by a tower of exponentials of fixed height, in the size of $\pi$ itself. This is a consequence of the following facts:}
\newcommand{\f}[1]{F#1}
\begin{enumerate}
	\item[F1.] reducing a cut at depth $i$ does not affect depth $j<i$;
	\item[F2.] cut-elimination does not increase the depth of \pn s;
	\item[F3.] reducing a cut at depth $i$ strictly decreases the size at depth $i$.
\end{enumerate}
\f 1 is true for all \MELL\ \pn s; \f 2 and \f 3 are consequences of the stratification property.

Now, the idea of \cite{Girard:LLL} is to eliminate cuts by operating at increasingly higher depths: if we have a \MultELL\ \pn\ of depth $d$, we start with a first ``round'' at depth $0$, which will eliminate all cuts at that depth in a finite amount of time because of \f 3; then, we proceed with a second round at depth~$1$, which, for the same reason, will eliminate all cuts at that depth, and will not create new cuts at depth $0$ because of \f 1; and we keep going on like this for all depths. By \f 2, this whole ``round by round'' procedure is guaranteed to terminate in at most $d+1$ rounds. After showing that the size of a \pn\ at the end of each round is at most $s^{s+1}<2^{2^s}$, where $s$ is the size of the \pn\ at the beginning of the round (this is analogous to \reflemma{SizeBoom}), one easily obtains an elementary bound in the size of the initial \pn, with the height of the tower of exponentials being at most twice the depth of the \pn\ itself. It is important to remark that the above argument makes no use of types: normalization in elementary size is possible even for untyped \MultELL\ \pn s.

\begin{figure}[t]
	\begin{center}\scalebox{0.9}{\input{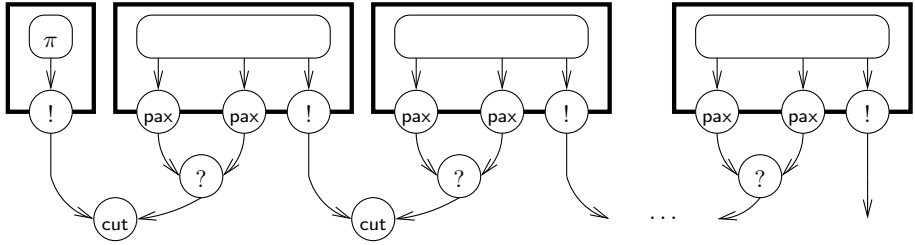}}\end{center}
	\caption{A chain of boxes causing an exponential blow-up in the size during cut-elimination.}
	\label{fig:BoxChain}
\end{figure}
\paragraph*{Box chains and light linear logic.} The reason for the superexponential blow-up in the size of \MultELL\ \pn s after each round can be understood intuitively by considering the ``chain'' of boxes of \reffig{BoxChain}. If the number of boxes with two \pax s in the chain is $n$, a simple calculation shows that there will be $2^n$ copies of $\pi$ when all cuts shown are reduced. In general, the \whynot\ links involved in a chain need to be binary; but their arity can be (very roughly) bounded by the size of the \pn\ containing the chain, and since the length of a chain can also be subjected to a similar bound, we end up obtaining the superexponential blow-up mentioned above.

If we want to moderate the increment of the size of \pn s under cut-elimination, by na\"\i vely looking at \reffig{BoxChain} we are led to think of a simple method: impose that boxes have at most one \pax. This actually turns out to work, and is the idea underlying Girard's \cite{Girard:LLL} definition of light linear logic. Unfortunately though, this restriction is quite heavy in terms of expressive power: in fact, while normalizable in polynomial time, \MultELL\ \pn s using boxes with at most one \pax\ are not able to compute all polytime functions. This is the original reason behind the introduction of the paragraph modality.

\begin{figure}[t]
	\begin{center}\scalebox{\scalefact}{\input{pargbox.tex}}\end{center}
	\caption{A $\S$-box.}
	\label{fig:PargBox}
\end{figure}
However, using the paragraph modality as we introduced it in \MELL\ is not compatible with the stratification property: the paragraph too must be linked to the depth, and in order to do so we must introduce a further kind of boxes, called \emph{$\S$-boxes} (\reffig{PargBox}). In presence of these boxes, the usual ones are called $\oc$-boxes, and the word ``box'' refers to any of the two kinds.
\begin{defn}[\MELLparg]
	\label{def:MELLparg}
	The \pps s and \ps s of \MELLparg\ are defined as in \refdef{ProofStructure}, with the following modifications on the requirements concerning boxes:
	\begin{itemize}
		\item[$a'$.] each \ofcourse\ link is the \pal{} of exactly one $\oc$-box;
		\item[$b$.] each \paxl\ link is in the border of exactly one box; 
		\item[$c$.] any two distinct boxes are either disjoint or included in one another;
		\item[$d$.] each \pargl\ link is in the border of exactly one $\S$-box.
	\end{itemize}

	The size of a \MELLparg\ \pps\ is defined just as in \refdef{DepthSize}, while the depth also takes into account $\S$-boxes, i.e., the depth of a link is the number of nested $\oc$- and $\S$-boxes containing it.

	The \pn s of \MELLparg\ are defined as in \refdef{ProofNet}, with $\S$-boxes being treated exactly as $\oc$-boxes.
\end{defn}

In terms of sequent calculus, a $\parg$-box corresponds to the following rule:
\begin{center}
	\mbox{\infer{\ts\wn\Gamma,\parg\Delta}{\ts\wn\Gamma,\Delta}}
\end{center}
After adapting \refdef{SeqProofStructure} to this rule, \refprop{Sequentialization} extends to \MELLparg.

To define cut-elimination inside \MELL$_\S$, one needs only to establish what the reduction of two $\S$-boxes looks like: informally, the two $\S$-boxes are ``merged'' into one, and the \cut\ link ``enters'' into this new $\S$-box. No detailed description is needed for our purposes; we refer the reader to~\cite{Mazza:LLAndPolytime}.

Multiplicative \LLL\ can be defined as a subsystem of \MELL$_\S$:
\begin{defn}[Multiplicative light linear logic]
	\label{def:LLL}
	\sloppy{Multiplicative light linear logic (\MLLL) is composed of all \MELLparg\ \pn s $\pi$ satisfying the following conditions:}
	\begin{description}
		\item[Depth-stratification:] Each exponential branch of $\pi$ crosses exactly one auxiliary port.
		\item[Lightness:] Each $\oc$-box of $\pi$ has at most one auxiliary port.
	\end{description}
\end{defn}
Observe that, in the depth-stratification condition, the \pax s of \mbox{$\S$-boxes} count just as the \pax s of $\oc$-boxes.

In the case of \MLLL, a round starting with a \pn\ of size $s$ can be shown to lead to a \pn\ of size at most $s^2$ (this is a special case of \reflemma{SizeBound}), so that the round-by-round procedure applied to a \pn\ of size $s$ and depth $d$ terminates with a \pn\ of size at most $s^{2^d}$.

\paragraph*{From size to time.} For the moment, we have only spoken of \emph{size} bounds to cut-elimination, whereas we started by claiming that \MultELL\ and \MLLL\ characterize \emph{time} complexity classes. The first step is transforming these size bounds into time bounds, which is done as follows. We consider the case of \MLLL, the case of \MultELL\ being analogous. Let $\pi$ be a \MLLL\ \pn\ of size $s$ and depth $d$. We know that we can eliminate all of its cuts in at most $d+1$ rounds, each operating on a \pn\ of size at most $s^{2^d}$. By \f 3, each round takes a linear number of steps in the size of the \pn\ from which the round itself starts; then, the round-by-round procedure for $\pi$ terminates in at most $(d+1)s^{2^d}$ steps.

Observe now that a single cut-elimination step can at most square the size of a \pn; then, with a reasonable representation of \pn s, we are able to simulate a cut-elimination step on a Turing machine with a polynomial cost, in the size of the \pn\ under reduction. Assuming that all \pn s during the reduction of $\pi$ have the maximum size possible, we have \mbox{$(d+1)s^{2^d}$} cut-elimination steps taking each $s^{2^{d+k}}$ Turing machine steps (where $k$ depends on the polynomial slowdown given by implementing cut-elimination on a Turing machine), which means that we can compute the result of the round-by-round procedure on $\pi$ in at most $(d+1)s^{2^{d+k}+2^d}$ Turing machine steps, which is polynomial in the size, and doubly-exponential in the depth. Similarly, computing the result of the round-by-round procedure for a \MultELL\ \pn\ takes a number of Turing machine steps which is elementary in the size, and hyperexponential in the depth.

\paragraph*{Representing functions.} To state precisely what it means for a logical system like \MultELL\ or \MLLL\ to characterize a complexity class, we first need to formulate a notion of \emph{representability} of functions from binary strings to binary strings. This is done by resorting to a formula (i.e., a type), which we may denote by $\Str$, such that there is an infinite number of \pn s of conclusion $\Str$, each representing a different binary string. It is very convenient at this point to operate within the intuitionistic subsystems of \MultELL\ and \MLLL, and to choose $\Str$ so that the \pn s of type $\Str$ correspond, via the computational interpretation discussed in \refsect{IntuitionLambda}, to the usual \lat s representing binary strings.

Then, we say that a function $f$ from binary strings to binary strings is representable in \MultELL\ or \MLLL\ just if there exists an intuitionistic \pn\ $\varphi$ of conclusions $\Str^\perp,\Str$ computing $f$ via cut-elimination, that is, $f(x)=y$ iff, whenever $\xi$ is the \pn\ representing $x$, the \pn\ $\varphi(\xi)$ obtained by cutting the conclusion (of type $\Str$) of $\xi$ to the dual conclusion (of type $\Str^\perp$) of $\varphi$ reduces to $\upsilon$, where $\upsilon$ is the \pn\ representing $y$. (Actually, it is necessary to allow representations of functions to be more generally of conclusions $\Str^\perp,\Str'$, where $\Str'$ is the formula $\Str$ with a number of suitable modalities prepended to it; but this is not essential at this level of detail).

\paragraph*{Characterizing complexity classes.} We say that a logical system characterizes a complexity class $\cC$ when $f\in\cC$ iff $f$ is representable in the logical system itself. The forward implication is usually called the \emph{completeness} of the system, while the backward implication is its \emph{soundness}.

Proving the completeness of \MultELL\ and \MLLL\ with respect to \FE\ and \FP, respectively, is a sort of (quite difficult) programming exercise, which is carried on with varying degrees of detail in \cite{Girard:LLL}, \cite{Roversi:Completeness}, \cite{DanosJoinet:ELL}, and \cite{MairsonTerui:CEP}; we shall not discuss this here.

On the other hand, the soundness of these two systems is a consequence of the results mentioned above, plus the following crucial remark: \emph{all \pn s of type $\Str$ have constant depth $1$, and size linear in the length of the string they represent}. Thanks to this, we see that if $\varphi$ is a \pn\ of \MLLL\ of size $s$ and depth $d$ representing the function $f$, and if $\xi$ represents the string $x$, then computing the representation of $f(x)$ can be done by applying the round-by-round cut-elimination procedure to the \pn\ $\varphi(\xi)$, whose size is $c_1|x|+c_2+s$ (where $c_1$ and $c_2$ are suitable constants), and whose depth is $\max(d,1)$, which \emph{does not depend on $x$}, but solely on $\varphi$, and thus, ultimately, on~$f$. Hence, $f(x)$ can be computed on a Turing machine in time $\mathcal O(P(|x|))$, where $P$ is a polynomial whose degree depends on~$f$. We therefore have $f\in\mathbf{FP}$. Similarly, one can prove that if $f$ is representable in \MultELL, then $f\in\mathbf{EF}$.

\section{Linear Logic by Levels}
\label{sect:LLlev}

\subsection{Indexings}
In \MELL\ \pn s there is an asymmetry between the behavior of the two kinds of exponential links (\ofcourse\ and \whynot) with respect to the depth. More precisely, let us say that a link $l$ is ``above'' an \ofcourse\ link $o$ if one of the conclusions of $l$ is the premise of $o$, and, similarly, let us say that $l$ is ``above'' a \whynot\ link $w$ if one of its conclusions is the premise of a \flatl\ link whose exponential branch (\refdef{ExpBranch}) ends in $w$. Then, we see that if a link $l$ is above an \ofcourse\ link $o$, we have $\Depth{l}=\Depth{o}+1$; on the contrary, if $l$ is above a \whynot\ link $w$, all we can say is that $\Depth{l}\geq\Depth{w}$.

The situation changes in \MultELL. In fact, the depth-stratification condition guarantees that the behavior is perfectly symmetric: if a link $l$ is above a \whynot\ link $w$, we have $\Depth{l}=\Depth{w}+1$. This is true also in \MLLL, and for \pargl\ links as well, because of $\S$-boxes (remember that, in \MLLL, the depth takes into account these boxes too).

The idea is then to take a \MELL\ \pn\ and to try assigning to its links an index which behaves as the depth would behave in elementary and light linear logic:
\begin{figure}[t]
	\begin{center}\scalebox{\scalefact}{\input{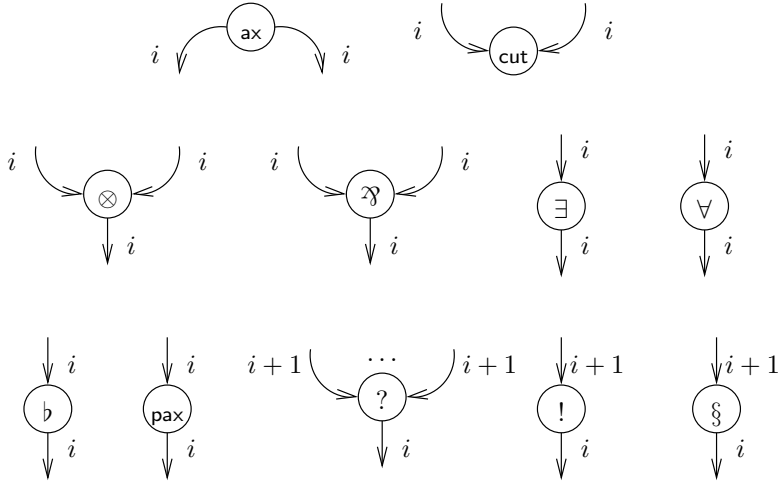}}\end{center}
	\caption{Constraints for indexing \MELL\ \pn s. Next to each edge we represent the integer assigned by the indexing; formulas are omitted, because irrelevant to the indexing.}
	\label{fig:Indexes}
\end{figure}
\begin{defn}[Indexing]\label{def:indexing}
	Let $\pi$ be a \MELL\ \ps. An \emph{indexing} for $\pi$ is a function $I$ from the edges of $\pi$ to $\Int$ satisfying the constraints given in \reffig{Indexes} and such that, for all conclusions $e,e'$ of $\pi$, $I(e)=I(e')$. An assignment satisfying the constraints of \reffig{Indexes} but not meeting the requirement on conclusions is said to be a \emph{weak indexing}.
\end{defn}
\begin{figure}[t]
	\begin{center}\scalebox{\scalefact}{\input{nonindexable.tex}}\end{center}
	\caption{A \MELL\ \pn\ admitting no (weak) indexing.}
	\label{fig:NonIndexable}
\end{figure}
Note that indexings do not use formulas in any way, so the notion can be applied to untyped \ps s without any change.

Not all \MELL\ \ps s admit an indexing. An example is the \pn\ in \reffig{NonIndexable}, which is the cut-free proof of the dereliction principle $\oc A\llinimp A$ (a key principle excluded in \ELL\ and \LLL). An analogous example is given by the two \pn s corresponding to the derivations mentioned in the proof of \refprop{PargIso}, i.e., the ones asserting the isomorphism between $A$ and $\S A$, although these do admit a weak indexing, contrarily to the \pn\ of \reffig{NonIndexable}.

Observe that weak indexings are transparent to connection: if $\pi_1,\pi_2$ are two \ps s admitting weak indexings $I_1,I_2$, respectively, then the \ps\ obtained by juxtaposing $\pi_1$ and $\pi_2$ admits as weak indexing the ``disjoint union'' of $I_1$ and $I_2$, which we denote by $I_1\uplus I_2$. Likewise, if $\pi$ is \ps\ whose connected components are $\pi_1,\ldots,\pi_n$, every (weak) indexing of $\pi$ can be written as $\biguplus I_k$, where $I_k$ is a (weak) indexing for $\pi_k$, for all $1\leq k\leq n$. We use this fact to state the following:
\begin{prop}[Rigidity]
	\label{prop:Rigidity}
	Let $\pi$ be a \MELL\ \ps\ whose connected components are $\pi_1,\ldots,\pi_n$, and let $I=\biguplus I_k$ be a (weak) indexing for $\pi$. Then, for all $p_1,\ldots,p_n\in\Int$, $\biguplus I_k+p_k$ is also a (weak) indexing for $\pi$. Conversely, given another (weak) indexing $I'$ for $\pi$, there exist $p_1,\ldots,p_n\in\Int$ such that $I'=\biguplus I_k+p_k$.
\end{prop}
\begin{pf}
	The first implication is trivial, so let us concentrate on the second. Let $I,I'$ be two (weak) indexings for $\pi$, and set, for each edge $e$ of $\pi$, $\Delta(e)=I(e)-I'(e)$. Now, observing \reffig{Indexes}, we see that differences in indexing propagate across any path in $\pi$; more precisely, whenever $e_1,e_2$ are both conclusions, both premises, or one conclusion and one premise of a link of $\pi$, then $\Delta(e_1)=\Delta(e_2)$. Hence, for any two edges $e,e'$ in the same connected component of $\pi$, we have $\Delta(e)=\Delta(e')$, which is enough to prove the result.\qed
\end{pf}

The following is a simple corollary of the first part of \refprop{Rigidity}:
\begin{prop}[Composition]
	Let $\pi,\pi'$ be two \pn s of resp.\ conclusions $\Gamma,A$ and $\Delta,A^\perp$, and let $\pi''$ be the \pn\ obtained by adding a cut link whose premises are the conclusions of $\pi$ and $\pi'$ labelled resp.\ by $A$ and $A^\perp$. Then, if $\pi$ and $\pi'$ both admit an indexing, so does $\pi''$.
\end{prop}

As a simple case-by-case inspection shows, indexings also have the fundamental property of being preserved under cut-elimination:
\begin{prop}[Stability]
	\label{prop:CutElimStab}
	Let $\pi$ be a \MELL\ \pn\ such that $\pi\onered\pi'$. Then, if there exists an indexing for $\pi$, there exists an indexing for $\pi'$ as well. More precisely, if $I$ is an indexing for $\pi$, there exists an indexing $I'$ of $\pi'$ such that, if $e,e'$ are conclusions of two links $l,l'$ of resp.\ $\pi,\pi'$ such that $l'$ is a residue of $l$, then $I'(e')=I(e)$. In other words, $I'$ is ``the same'' indexing as $I$, modulo the erasures/duplications possibly induced by the cut-elimination step.
\end{prop}

We can therefore give the following definition:
\begin{defn}[Multiplicative linear logic by levels]
	\sloppy{Multiplicative linear logic by levels (\MELLlev) is the logical system defined by taking all \MELL\ \pn s admitting an indexing.}
\end{defn}

The fact that an \MELLlev\ \pn\ has several (in fact, an infinity of) indexings may seem inconvenient; however, \refprop{Rigidity} settles this problem, by giving us a way to choose a \emph{canonical indexing}:
\begin{defn}[Canonical indexing]
	\label{def:CanInd}
	Let $\pi$ be an \MELLlev\ \pn, and let $I$ be an indexing for $\pi$. We say that $I$ is \emph{canonical} if each connected component of $\pi$ has an edge $e_0$ such that $I(e_0)=0$, and $I(e)\geq 0$ for all edges $e$ of $\pi$.
\end{defn}
\begin{prop}
	Every \MELLlev\ \pn\ admits a unique canonical indexing.
\end{prop}
\begin{pf}
	Let $\pi$ be an \MELLlev\ \pn, let $\pi_1,\ldots,\pi_n$ be the connected components of $\pi$, and let $k$ range over $\{1,\ldots,n\}$. By definition, there exists an indexing $\biguplus I_k$ for $\pi$, where $I_k$ is an indexing for $\pi_k$. Let $m_k=\min_{e} I_k(e)$, where $e$ ranges over the edges of $\pi_k$. Then, by \refprop{Rigidity}, $\biguplus I_k-m_k$ is still an indexing for $\pi$, which is clearly canonical. Suppose now there exist two canonical indexes $I=\biguplus I_k$ and $I'=\biguplus I_k'$ for $\pi$. By the fact that $I$ and $I'$ are canonical, we know that for all $k$ there exist $e_k,e_k'$ in $\pi_k$ such that $I(e_k)=I'(e_k')=0$. By \refprop{Rigidity}, we also know that there exists $p_k\in\Int$ such that $I_k'=I_k+p_k$. Suppose $p_k>0$; then, we would have $I(e_k')<0$. On the other hand, if $p_k<0$, we would have $I'(e_k)<0$. In both cases, we would be in contradiction with the fact that $I$ and $I'$ are canonical, hence we must have $p_k=0$, and $I=I'$.\qed
\end{pf}
\begin{defn}[Level]
	\label{def:Level}
	Let $\pi$ be an \MELLlev\ \pn, and let $I_0$ be its canonical indexing. The \emph{level} of $\pi$, denoted by $\Level{\pi}$, is the maximum integer assigned by $I_0$ to the edges of $\pi$. If $l$ is a link of $\pi$ of conclusion $e$ (or of conclusions $e_1,e_2$ in the case of an \axiom\ link), and if $\cB$ is a box of $\pi$ whose principal port has conclusion $e'$, we say that the level of $l$, denoted by $\Level l$, is $I_0(e)$ (or $I_0(e_1)=I_0(e_2)$ in the case of an axiom), and that the level of $\cB$, denoted by $\Level\cB$, is $I_0(e')$.
\end{defn}

From now on, when we speak of an \MELLlev\ \pn\ $\pi$, we shall always refer to its canonical indexing. The reader may wonder why we did not use $\Nat$ instead of $\Int$ as the range of our indexes in the first place; we simply believe $\Int$ to be a more natural choice, as the set of indexes need not be well-founded. Moreover, using $\Nat$ would be awkward in the sequent calculus formulation of \MELLlev\ (cf.\ \reftab{MELLlevSeq} below): it would force to impose a restriction on exponential rules, an unnecessary complication. Remark also that \refprop{Rigidity} shows that the set of (weak) indexings of a proof net with $n$ connected components forms an affine space over the module $\Int^n$ (in the case of indexings, all components having a conclusion must be considered as one connected component); indeed, the canonical indexing is just a way of fixing an ``origin'' for such affine space. This nice algebraic structure, which we shall not investigate more in this work, is a further motivation to the use of relative integers instead of natural integers.

Recall that levels are conceived to behave like depths in \MultELL; then, it is not surprising that \MultELL\ is exactly the (proper) subsystem of \MELLlev\ in which levels and depths coincide:
\begin{prop}
	\label{prop:ELL}
	Let $\pi$ be a \MELL\ \pn. Then, $\pi$ is in \MultELL\ iff $\pi$ is in \MELLlev\ and, for every link $l$ of $\pi$ whose conclusion is not a \struct, we have $\Level l=\Depth l$.
\end{prop}

Note that \MultELL\ is not only a proper subsystem of \MELLlev\ at level of proofs, but also at the level of provability. For instance, we invite the reader to check that the formula $\oc{(\oc A\ltens B)}\llinimp\oc{\oc A}\ltens\wn B$ is provable in \MELLlev, but not in \MultELL.

Now to help relating \pn s to the intuitions coming from the \lac, we give an example of a \lat\ and a corresponding \pn\ of \MELLlev. The following term is the Church representation of the binary list $101$, and its syntactic tree is given in \reffig{Exempleterme}:
\begin{figure}[t]
	\begin{center}\scalebox{\scalefact}{\input{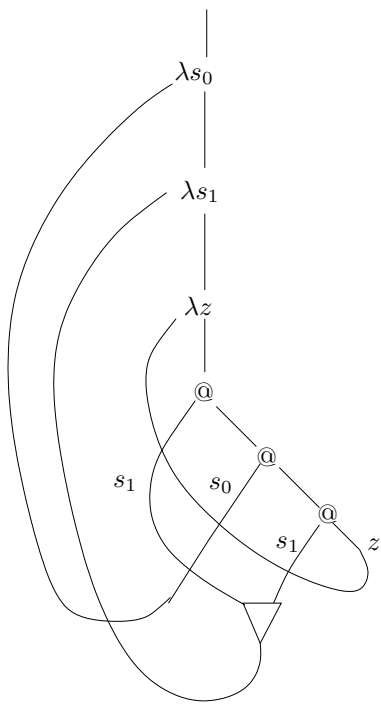}}\end{center}
	\caption{Syntactic tree for the $\lambda$-term $t_{101}$.}
	\label{fig:Exempleterme}
\end{figure}
$$ t_{101}=\lambda s_0.\lambda s_1. \lambda z. (s_1 \;  (s_0 \;  (s_1 \; z))).$$
An \MELLlev\ \pn\ corresponding to this term, according to \refprop{ComputSem}, is given in \reffig{Exemplereseau}. Note that nodes $\lambda$ (resp. $@$) of the syntactic tree correspond to nodes $\lpar$ (resp. $\ltens$) of the \pn.
\begin{figure}[t]
	\begin{center}\scalebox{\scalefact}{\input{PatExemplereseau.tex}}\end{center}
	\caption{An  \MELLlev\  proof-net corresponding to $t_{101}$.}
	\label{fig:Exemplereseau}
\end{figure}

\subsection{Light linear logic by levels}
\label{sect:LLLlev}
Chains of boxes like that of \reffig{BoxChain} may be built in \MELLlev, so there is no hope of finding sub-exponential bounds for the size of \MELLlev\ \pn s under cut-elimination. We then follow the same idea as light linear logic:
\begin{defn}[Multiplicative light linear logic by levels]
	\label{def:MLLLlev}
	\sloppy{Multiplicative light linear logic by levels (\MLLLlev) is the logical system composed of all \MELLlev\ \pn s $\pi$ satisfying the following conditions:}
	\begin{description}
		\item[(Weak) Depth-stratification:] Each exponential branch (\refdef{ExpBranch}) of $\pi$ crosses at most one auxiliary port.
		\item[Lightness:] Each box of $\pi$ has at most one auxiliary port.
	\end{description}
\end{defn}

It is not hard to see that \MLLLlev\ is stable under cut-elimination, i.e., that a suitable version of \refprop{CutElimStab} holds. Indeed, the depth-stratification condition is needed precisely for that purpose: in its absence, one can find an \MELLlev\ \pn\ satisfying the lightness condition which reduces to a \pn\ no longer satisfying it.

As expected, \MLLLlev\ is related to \MLLL. To see how, we consider the forgetful embedding of \MLLL\ into \MELL\ which simply removes paragraph boxes, retaining only the corresponding paragraph links (recall that our definition of \MELL\ includes the paragraph modality). Observe that this embedding is compatible with cut-elimination: if $\pi_1\onered\pi_2$, then $\pi_1^+\onered\pi_2^+$ (see \cite{Mazza:LLAndPolytime} for the details on cut-elimination with $\S$-boxes). We can then see \MLLL\ as a subsystem of \MLLLlev, in the following sense:
\begin{prop}
	\label{prop:LLL}
	Let $\pi$ be a \MLLL\ \pn, and let $\pi^+$ be its forgetful image in \MELL. Then, $\pi^+$ is in \MLLLlev\ and, for every link $l^+$ of $\pi^+$ whose conclusion is not a \struct\ and which corresponds to a link $l$ of $\pi$, we have $\Level{l^+}=\Depth{l}$ (we remind that in \MLLL\ \pn s the depth also takes into account paragraph boxes, see \refdef{MELLparg}).
\end{prop}

As already observed above, $\S A$ is not isomorphic to $A$ in \MELLlev\ (or \MLLLlev). However, it is not hard to check that in both systems the paragraph modality commutes with all connectives: for all $A,B$, $\S(A\ltens B)$, $\S\oc A$, and $\S\forall X.A$ are all provably isomorphic (in the same sense as that of \refprop{PargIso}) to $\S A\ltens\S B$, $\oc\S A$, and $\forall X.\S A$, respectively (and, by duality, similar isomorphisms hold for the connectives $\lpar$, $\wn$, and $\exists$).

None of the above isomorphisms holds in \LLL, and this is why it does not make much sense to establish a converse of \refprop{LLL}. We therefore obtained a system in which the paragraph modality, like \LLL, is not trivial, but, unlike \LLL, enjoys more flexible principles. In \refsect{Complexity} we shall see that \MELLlev\ and \MLLLlev\ have also interesting properties with respect to the complexity of their cut-elimination procedure.

\subsection{Linear logic by levels as a sequent calculus}
\label{sect:Seq}
It is possible to formulate \MELLlev\ and \MLLLlev\ as sequent calculi, which may be useful for having a clearer correspondence with \lat s, as in \refsect{IntuitionLambda}. In doing this, one immediately realizes that \emph{2-sequents}, rather than sequents, are the natural syntax for this purpose. Calculi for 2-sequents have been extensively studied by \cite{Masini:TwoSeq} and have been found to be quite useful for the proof-theory of modal logics. In particular, linear logic and its elementary and light variants can all be formulated as 2-sequent calculi~\citep{GuerriniMartiniMasini:TwoSeq}.

A \MELL\ \emph{2-sequent} $M$ is a function from $\Int$ to \MELL\ sequents such that $M(i)$ is the empty sequent for all but finitely many $i$. 2-sequents can be succinctly represented as standard sequents by decorating formulas with an integer index: $\ts A_1^{i_1},\ldots,A_n^{i_n}$ represents the 2-sequent $M$ such that \mbox{$M(i)=\mbox{$\ts\Gamma$}$}, where $\Gamma$ contains all and only the occurrences of formulas $A_j^{i_j}$ such that $i_j=i$.

\begin{table}[t]
	\TwoRules
	{\infer[\etq{Axiom}]{\ts A^{\perp i},A^i}{}}
	{\infer[\etq{Cut}]{\ts\Gamma,\Delta}{\ts\Gamma,A^i & \ts\Delta,A^{\perp i}}}
	\smallskip
	\TwoRules
	{\infer[\etq{Tensor}]{\ts\Gamma,\Delta,{A\ltens B}^i}{\ts\Gamma,A^i & \ts\Delta,B^i}}
	{\infer[\etq{Par}]{\ts\Gamma,{A\lpar B}^i}{\ts\Gamma,A^i,B^i}}
	\smallskip
	\TwoRules
	{\infer[\etq{For all ($X$ not free in $\Gamma$)}]{\ts\Gamma,{\forall X.A}^i}{\ts\Gamma,A^i}}
	{\infer[\etq{Exists}]{\ts\Gamma,{\exists X.A}^i}{\ts\Gamma,{A[B/X]}^i}}
	\smallskip
	\TwoRules
	{\infer[\etq{Promotion}]{\ts\wn\Gamma,\oc A^i}{\ts\wn\Gamma,A^{i+1}}}
	{\infer[\etq{Dereliction}]{\ts\Gamma,\wn A^i}{\ts\Gamma,A^{i+1}}}
	\smallskip
	\TwoRules
	{\infer[\etq{Weakening}]{\ts\Gamma,\wn A^i}{\ts\Gamma}}
	{\infer[\etq{Contraction}]{\ts\Gamma,\wn A^i}{\ts\Gamma,\wn A^i,\wn A^i}}
	\smallskip
	\OneRule
	{\infer[\etq{Paragraph}]{\ts\Gamma,{\parg A}^i}{\ts\Gamma,A^{i+1}}}
	\caption{The rules for \MELLlev\ 2-sequent calculus. Daimon and mix are omitted.}
	\label{tab:MELLlevSeq}
\end{table}
The 2-sequent calculus for \MELLlev\ is given in \reftab{MELLlevSeq}, where $\Gamma,\Delta$ stand for multisets of \MELL\ formulas decorated with an integer. The daimon and mix rules are omitted, because identical to those in \reftab{MELLSeq}.

We say that a derivation of $\ts\Gamma$ in the calculus of \reftab{MELLlevSeq} is \emph{proper} if all the formulas in $\Gamma$ have the same index, i.e., the derived 2-sequent is indeed a sequent; moreover, we say that a \emph{weak} \MELLlev\ \ps\ is a \ps\ admitting a weak indexing. By \refprop{Sequentialization}, it is more or less evident that a sequentializable weak \MELLlev\ \ps\ is a weak \MELLlev\ \pn. Hence, we see that \MELLlev\ \pn s exactly correspond to the proper derivations of the calculus of \reftab{MELLlevSeq}.

We remark that the calculus of \reftab{MELLlevSeq} is very similar to Guerrini, Martini, and Masini's \TwoELL~\citep{GuerriniMartiniMasini:TwoSeq}, without additive connectives: the two calculi differ in the formulation of the promotion rule (whose context, in \TwoELL, need not be of the form $\wn\Gamma$) and in a series of constraints imposed on some rules of \TwoELL\ (in particular on promotion). In their work, the authors show that cut-free provability in \TwoELL\ coincides with provability in \ELL, leaving open the question of whether \TwoELL\ satisfies cut-elimination. All the constraints of the multiplicative fragment of \TwoELL\ are removed in our calculus, and in fact \MELLlev\ is a proper extension of \MultELL, both in terms of proofs and provability---preserving, however, its complexity properties, as we shall see below.

The system \MLLLlev\ is obtained in sequent calculus by replacing the promotion rule with the following one:
\OneRule{\infer[\etq{Light promotion}]{\ts\wn B^j,\oc A^i}{\ts B^{j+1},A^{i+1}}}
where the formula $B$ may not be present.

\section{Complexity Bounds}
\label{sect:Complexity}
To establish the complexity bounds for \MELLlev\ and \MLLLlev, we shall try to adapt the arguments originally given by \cite{Girard:LLL} for \ELL\ and \LLL. Let us then go back to \refsect{ELLandLLL} and consider again the three facts about cut-elimination in \MultELL\ which are at the base of its elementary size bound:
\begin{enumerate}
	\item[F1.] reducing a cut at depth $i$ does not affect depth $j<i$;
	\item[F2.] cut-elimination does not increase the depth of \pn s;
	\item[F3.] reducing a cut at depth $i$ strictly decreases the size at depth $i$.
\end{enumerate}
We know that \f 1 is true in general in \MELL, and hence in \MELLlev\ too; it is not hard to see that \f 2 and \f 3 instead fail altogether in \MELLlev\ and \MLLLlev. Nevertheless, in the light of Propositions~\ref{prop:ELL}~and~\ref{prop:LLL}, we may expect those facts to hold in our systems provided we replace the word ``depth'' with ``level''. Indeed, this works for \f 2: 
\begin{lem}
	\label{lemma:Level}
	Let $\pi$ be an \MELLlev\ \pn\ such that $\pi\onered\pi'$. Then, \mbox{$\Level{\pi'}\leq\Level\pi$}.
\end{lem}

\begin{figure}[t]
	\begin{center}\scalebox{0.65}{\input{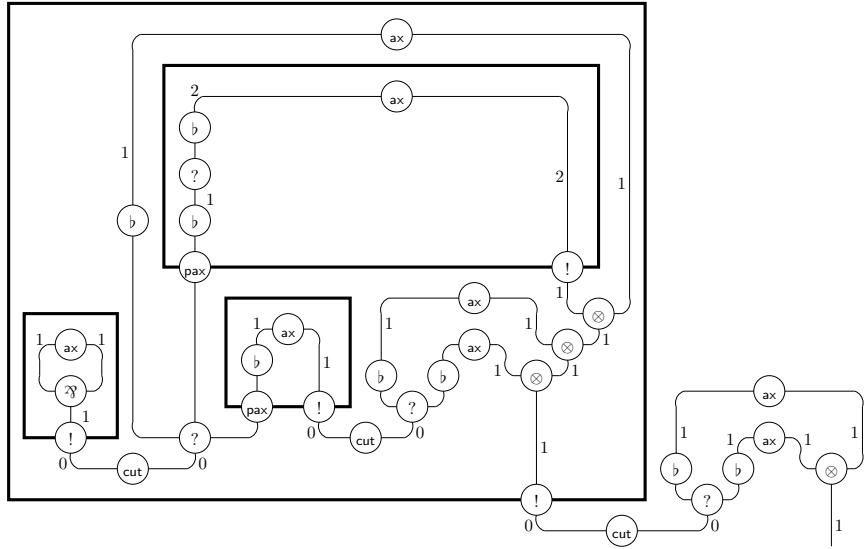}}\end{center}
	\caption{An example of nested boxes of identical level (much smaller examples exist; we gave this one because we shall re-use it later on for different purposes).}
	\label{fig:NestedLevel}
\end{figure}
On the contrary, the ``level-wise'' versions of \f 1 and \f 3 fail for \MELLlev\ and \MLLLlev, because a box of level $i$ may contain links of \emph{any} level, in particular $i$ itself. \reffig{NestedLevel} gives an example of this: reducing a cut at level $i$ ($i=0$ in this case) may duplicate cuts at the same level. Therefore, a straightforward adaptation of Girard's ``round-by-round'' procedure, which trades depths for levels, will not work. There is a workaround though: in fact, there are cuts for which the level-wise version of \f 3 holds, and for which the failure of \f 1 is harmless; our solution will consist in showing that these can be reduced first.

\subsection{Termination}
First of all, we prove that reduction of \MELLlev\ \pn s always terminates, even in the untyped version of the system. From this moment on, that is, for the rest of \refsect{Complexity}, by ``\MELL\ \pn'' we shall mean ``untyped \MELL\ \pn'', and by ``\MELLlev\ (resp.\ \MLLLlev) \pn'' we shall mean ``untyped \MELL\ \pn\ admitting an indexing (resp.\ admitting an indexing and satisfying the structural conditions of \refdef{MLLLlev})''.
\begin{defn}[Isolevel tree]
	Let $\pi$ be a \MELL\ \pn, and let $e$ be an edge of $\pi$ which is the conclusion of a link $l$ different from \flatl\ or \paxl. The \emph{isolevel tree} of $e$ is defined by induction as follows:
	\begin{itemize}
		\item if $l$ is an \axiom, \whynot, \ofcourse, or \pargl\ link, then the isolevel tree of $e$ consists of the link $l$ alone;
		\item otherwise, let $e_1,\ldots,e_k$ (with $k\in\{1,2\}$) be the premises of $l$; then, the isolevel tree of $e$ is the tree whose root is $l$ and whose immediate subtrees are the isolevel trees of $e_1,\ldots,e_k$.
	\end{itemize}
\end{defn}
\begin{defn}[Complexity of reducible cuts]
	Let $\pi$ be a \MELL\ \pn, and let $c$ be a reducible \cut\ link of $\pi$, whose premises are $e_1,e_2$. The \emph{complexity} of $c$, denoted by $\Compl c$, is the sum of the number of nodes contained in the isolevel trees of $e_1$ and $e_2$. (Note that the isolevel trees of $e_1,e_2$ are always defined because the premises of a \cut\ can never be conclusions of \flatl\ or \paxl\ links).
\end{defn}
\begin{defn}[Weight of an \MELLlev\ \pn]
	\label{def:Wght}
	Let $\pi$ be an \MELLlev\ \pn\ of level $l$. If $k\in\Int$, we denote by $\Cuts{k}{\pi}$ the set of reducible \cut\ links of $\pi$ at level $k$. The \emph{weight} of $\pi$, denoted by $\Wght\pi$, is the function from $\Nat$ to $\Nat$ defined as follows:
	\begin{displaymath}
		\Wght{\pi}(i)=\sum_{c\in\Cuts{l-i}{\pi}}\Compl c.
	\end{displaymath}
\end{defn}

Note that, if $\pi$ has level $l$, then for all $i>l$, we have $\Wght{\pi}(i)=0$. Weights are therefore almost everywhere null, and the set of all weights can be well-ordered so as to be isomorphic to $\omega^\omega$.

We recall that, concretely, this order is a variant of the lexicographical order, and is defined as follows. Let $\alpha,\beta$ be two almost-everywhere-null functions from $\Nat$ to $\Nat$. We put $C_{\alpha,\beta}=\{i\in\Nat~;~\alpha(i)\neq\beta(i)\}$. Observe that $C_{\alpha,\beta}$ is finite, because $\alpha$ and $\beta$ are almost everywhere null. Moreover, $C_{\alpha,\beta}$ is non-empty iff $\alpha\neq\beta$; in this case, let $m=\max C_{\alpha,\beta}$, and we set $\alpha<\beta$ iff $\alpha(m)<\beta(m)$.

So for all $\pi$, $\Wght{\pi}$ can be seen as an ordinal strictly smaller than $\omega^\omega$. Our cut-elimination proof will simply show that, whenever an \MELLlev\ \pn\ $\pi$ is not normal, there always exists $\pi'$ such that $\pi\onered\pi'$ and $\Wght{\pi'}<\Wght{\pi}$.

Below, we say that a \flatl\ link $b$ is \emph{above} a \whynot\ link $w$ iff the exponential branch of $b$ ends in $w$.
\begin{defn}[Contractive order]
	Let $\pi$ be an \MELLlev\ \pn, and let $\cB,\cC$ be two boxes of $\pi$. We write $\cB\immctrprec\cC$ iff $\cB$ and $\cC$ are at the same level, $\cB$ is cut with a \whynot\ link $w$, and $\cC$ contains a \flatl\ link above $w$. We denote by $\ctrprec$ the reflexive-transitive closure of $\immctrprec$.
\end{defn}
\begin{lem}
	\label{lemma:POrd}
	The relation $\ctrprec$ is a partial order.
\end{lem}
\begin{pf}
	Suppose there is a cycle in $\immctrprec$, i.e., there exist $n\geq 1$ different boxes $\cB_1,\ldots,\cB_n$ such that $\cB_1\immctrprec\cdots\immctrprec\cB_n\immctrprec\cB_1$. We say that such a cycle has a \emph{lump} iff there exist $i\neq j$ such that $\cB_i\immctrprec\cB_j$ and $\cB_i$ is contained in $\cB_j$. Let $k$ be the number of lumps in the cycle; we shall prove a contradiction by induction on $k$. If $k=0$, then all boxes are disjoint. In this case, it is easy to build, by induction on $n$, a cyclic switching of $\pi$ (or of the contents of the minimal box containing the whole chain), which is impossible, since $\pi$ is supposed to be a \pn. If $k>0$, let $\cB_i,\cB_j$ be a pair of boxes inducing a lump. Since we have a cycle, there certainly exists $p$ such that $\cB_p\immctrprec\cB_i$. If $p=j$, then there is obviously a cyclic switching around $\cB_j$, yielding again a contradiction. Otherwise, by definition, $\cB_p\immctrprec\cB_i$ means that there is a \flatl\ link inside $\cB_i$ which is above the \whynot\ link to which $\cB_p$ is cut. But $\cB_i$ is contained in $\cB_j$, so this \flatl\ link is also in $\cB_j$, which means that $\cB_p\immctrprec\cB_j$ as well. Independently of whether $\cB_p$ is included in $\cB_j$ or not, the cycle obtained by removing $\cB_i$ from the original one necessarily has $k-1$ lumps, and the induction hypothesis applies. Therefore, $\immctrprec$ is acyclic, and its reflexive-transitive closure is a partial order.\qed
\end{pf}

In the following, we deem a \cut\ link \emph{contractive} iff its premises are the conclusions of an \ofcourse\ link and a \whynot\ link of arity strictly greater than zero. All other reducible \cut\ links are called \emph{non-contractive}.
\begin{defn}[Cut order]
	\label{def:CutOrd}
	Let $\pi$ be an \MELLlev\ \pn, and let $\AllCuts{\pi}$ be the set of reducible \cut\ links of $\pi$. We turn $\AllCuts{\pi}$ into a partially ordered set by posing, for $c,c'\in\AllCuts{\pi}$, $c\leq c'$ iff one of the following holds:
	\begin{itemize}
		\item $\Level{c}<\Level{c'}$;
		\item $c$ is non-contractive and $c'$ is contractive;
		\item $c$ and $c'$ are both contractive, involving resp.\ the boxes $\cB$ and $\cB'$, and $\cB\ctrprec\cB'$.
	\end{itemize}
\end{defn}
That the above relation is indeed a partial order follows easily from the definition and \reflemma{POrd}.

The weak normalization of untyped \MELLlev\ is a trivial corollary of the following result, as anticipated above:
\begin{lem}
	\label{lemma:WN}
	Let $\pi$ be an \MELLlev\ \pn\ which is not normal. Then, there exists $\pi'$ such that $\pi\onered\pi'$ and $\Wght{\pi'}<\Wght{\pi}$.
\end{lem}
\begin{pf}
	By hypothesis, $\AllCuts{\pi}\neq\emptyset$; of course $\AllCuts{\pi}$ is also finite, so there is at least one minimal element w.r.t.\ the cut order. Take any one of them (call it $c$), and reduce it, obtaining $\pi'$. Let $M$ (resp.\ $M'$) be the maximum $k$ such that $\Wght{\pi}(k)>0$ (resp.\ $\Wght{\pi'}(k)>0$). First of all, using \reflemma{Level}, we have that $\Level{\pi'}\leq\Level\pi$ and $M'\leq M$. If any of the two inequalities is strict, we immediately have $\Wght{\pi'}<\Wght\pi$. Therefore, we may assume $\Level{\pi'}=\Level\pi=l$ and $M'=M$. By the minimality hypothesis, we see that the level of $c$ must be $i=l-M$, and that $\pi$ contains no reducible cut at level $j<i$. This implies that, whatever happens in reducing $c$, $\Wght{\pi'}(n)=\Wght{\pi}(n)=0$ for all $n>M$, so it is enough to check that something decreases at level $i$, i.e., that $\Wght{\pi'}(M)<\Wght{\pi}(M)$. The proof now splits into five cases, depending on the nature of $c$. If $c$ is not an exponential cut, or if it is a weakening cut, we leave it to the reader to verify that the condition holds.

	So let $c$ be contractive, and let $\cB$ be the box involved. We claim that the content of $\cB$ contains no reducible \cut\ links at level $i$. As a matter of fact, suppose for the sake of contradiction that $\cB$ contains a reducible cut $c'$ of level $i$ (which is necessarily different from $c$). Because of the second clause of \refdef{CutOrd}, $c'$ must be contractive, otherwise we would contradict the minimality of $c$. But in this case, let $\cB'$ and $w$ be resp.\ the box and the \whynot\ link involved in $c'$. Since $c'$ is contractive, there is at least one \flatl\ link above $w$, which entails $\cB'\ctrprec\cB$; by the third clause of \refdef{CutOrd}, we would thus obtain a second, definitive contradiction.

	Now that we know that $\cB$ is normal at level $i$, it is not hard to verify that the thesis holds: $\pi'$ contains at least one copy of the content of $\cB$, but none of these copies contributes to the value of $\Wght{\pi'}(M)$. Moreover, the new cuts contained in $\pi'$ are all at level $i+1$, whereas one reducible cut at level $i$ ($c$ itself) has disappeared. Therefore, $\Wght{\pi'}(M)<\Wght{\pi}(M)$, as desired.\qed
\end{pf}

\begin{prop}[Untyped weak normalization]
	\label{prop:WN}
	Untyped \MELLlev\ \pn s are weakly normalizable.
\end{prop}
\begin{pf}
	By transfinite induction up to $\omega^\omega$. Let $\beta<\omega^\omega$, and suppose that for all $\alpha<\beta$, $\Wght{\pi}=\alpha$ implies that $\pi$ is weakly normalizable. Take a \pn\ $\pi$ such that $\Wght{\pi}=\beta$; $\pi$ is either normal, hence weakly normalizable, or, by \reflemma{WN} and by the above induction hypothesis, it reduces to a weakly normalizable \pn. But any \pn\ reducing to a weakly normalizable \pn\ is also weakly normalizable.\qed
\end{pf}

\subsection{Elementary bound for \MELLlev}
From now on, we shall only consider the cut-elimination procedure given by the proof of \reflemma{WN}, i.e., the one reducing only minimal cuts in the cut order. More concretely, given an \MELLlev\ \pn\ $\pi$, this procedure chooses a cut to be reduced in the following way:
\begin{enumerate}
	\item find the lowest level at which reducible cuts are present in $\pi$, say $i$;
	\item if non-contractive cuts are present at level $i$, choose any of them and reduce it;
	\item if only contractive cuts are left, chose one involving a minimal box in the contractive order.
\end{enumerate}
This is nothing but Girard's ``round by round'' procedure, modulo two modifications: we use levels instead of depths, and we are more restrictive on which contractive cuts can be reduced (in Girard's procedure for \MLLL, \emph{any} contractive cut may be reduced once all non-contractive cuts at the same depth are reduced). This last point is strictly technical: it is required because of configurations such as the one shown in \reffig{NestedLevel}, as discussed above. What is really fundamental is the shift from depth to level, which is indeed the key novelty of our work.

Let us start with a few useful definitions:
\begin{defn}
	Let $\pi$ be an \MELLlev\ \pn.
	\begin{enumerate}
		\item The \emph{size of level $i$} of $\pi$, denoted by $\ISize{i}{\pi}$, is the number of links at level $i$ of $\pi$ different from \pax s.
		\item $\pi$ is \emph{$i$-normal} iff it contains no reducible \cut\ link at all levels $j\leq i$.
		\item $\pi$ is \emph{$i$-contractive} iff it is $(i-1)$-normal and contains only contractive \cut\ links at level $i$.
	\end{enumerate}
\end{defn}
\begin{lem}
	\label{lemma:Linear}
	Let $\pi$ be an $(i-1)$-normal \pn. Then, the round-by-round procedure reaches an $i$-normal \pn\ in at most $\ISize{i}{\pi}$ steps.
\end{lem}
\begin{pf}
	Let $\pi=\pi_0\onered\pi_1\onered\cdots\onered\pi_n$ be reduction sequence generated by our procedure, with $\pi_n$ $i$-normal. By what we have seen in the proof of \reflemma{WN}, if we put $M=\Level\pi-i$, we have that $\Wght{\pi_{j+1}}(M)<\Wght{\pi_j}(M)$ for all $0\leq j\leq n-1$. Therefore, $n\leq\Wght\pi(M)$. But by definition $\Wght{\pi}(M)\leq\ISize{i}{\pi}$, hence the thesis.\qed
\end{pf}
Below, we use the notation $2_k^n$ with the following meaning: for all $n$, $2_0^n=n$, and $2_{k+1}^n=2^{2_k^n}$.
\begin{lem}
	\label{lemma:SizeBoom}
	Let $\pi$ be an $i$-contractive \pn, such that $\pi\red\pi'$ under the round-by-round procedure, with $\pi'$ $i$-normal. Then, $\Size{\pi'}\leq 2_2^{\Size\pi}$.
\end{lem}
\begin{pf}
	In the proof, we shall say that the arity of a contractive \cut\ link $c$ is the arity of the \whynot\ link whose conclusion is premise of $c$. Let $\pi_0$ be an $i$-contractive \pn, such that $\pi_0\onered\pi_1$ by reducing a minimal cut $c$ at level~$i$. We have that, for all $k\neq i$, $\ISize{k}{\pi_0}=B_k+C_k$, while $\ISize{i}{\pi_0}=B_i+C_i+3$, where $B_k$ is the size of level $k$ of the content of the box $\cB$ whose principal port's conclusion is premise of $c$, and $C_k$ is a suitable non-negative integer. It is enough to inspect \reffig{ExpStep} to see that, if the arity of $c$ is $A$, we have $\ISize{k}{\pi_1}=AB_k+C_k$, for all $k$. Now, since the step is contractive, $A\geq 1$, so that $\ISize{k}{\pi_1}\leq A(B+C)=A\ISize{k}{\pi_0}$.

	We now make the following claims:
	\begin{enumerate}
		\item $\pi_1$ is $i$-contractive;
		\item if $c_1$ is \cut\ link of $\pi_1$ at level $i$, and $c_0$ is its lift in $\pi_0$, then the arities of $c_0$ and $c_1$ coincide.
	\end{enumerate}
	The first fact can be checked by simply looking at \reffig{ExpStep}. For what concerns the second, let $w_0,\cB_0$ and $w_1,\cB_1$ be resp.\ the \whynot\ link and box cut by resp.\ $c_0$ and $c_1$. Note that, by hypothesis, $w_0$ and $\cB_0$ are the lifts of resp.\ $w_1$ and $\cB_1$. Now suppose, for the sake of contradiction, that the arity of $w_1$ is different than that of $w_0$. Another simple inspection of \reffig{ExpStep} shows that this may be the case only if an exponential branch of $\pi_0$ ending in $w_0$ crosses the border of $\cB$ (the box involved in the reduction leading from $\pi_0$ to $\pi_1$). But if it is so, then there is a \flatl\ link above $w_0$ which is inside $\cB$, which implies that $\cB_0\ctrprec\cB$. By \refdef{CutOrd}, we have $c_0<c$, contradicting the minimality of $c$. Therefore, the maximum arity of all cuts of $\pi_1$ at level $i$ cannot exceed the maximum arity of all cuts of $\pi_0$ at level $i$.

	Let now $\pi=\pi_0\onered\cdots\onered\pi_n=\pi'$ be the reduction sequence generated by the round-by-round procedure. If $A_1,\ldots,A_n$ are the arities of the \cut\ links reduced at each step, we have, for all $k$,
	$$\ISize{k}{\pi'}\leq\ISize{k}{\pi}\prod_{j=1}^nA_j.$$
	But, by the above claim, each $A_j$ cannot be greater than the greatest arity of \whynot\ links present in $\pi$. This is of course bounded by $\ISize{i+1}{\pi}$ (a contraction of arity $A$ at level $i$ needs the presence of $A$ \flatl\ links at level $i+1$), so we can conclude that
	$$\ISize{k}{\pi'}\leq\ISize{k}{\pi}\ISize{i+1}{\pi}^n\leq\ISize{k}{\pi}\ISize{i+1}{\pi}^{\ISize{i}{\pi}},$$
	where we have used \reflemma{Linear}, which tells us that $n\leq\ISize{i}{\pi}$. Now, if put $l=\Level{\pi'}=\Level\pi$, we have
	$$\Size{\pi'}=\sum_{k=0}^l\ISize{k}{\pi'}\leq\sum_{k=0}^l\ISize{k}{\pi}\ISize{i+1}{\pi}^{\ISize{i}{\pi}}=\Size\pi\ISize{i+1}{\pi}^{\ISize{i}{\pi}}\leq{\Size\pi}^{\Size\pi+1}\leq 2^{2^{\Size\pi}},$$
	as stated in our thesis.\qed
\end{pf}
\begin{thm}[Elementary bound for \MELLlev]
	\label{th:ElemBound}
	Let $\pi$ be an \MELLlev\ \pn\ of size $s$ and level $l$. Then, the round-by-round procedure reaches a normal form in at most $(l+1)2_{2l}^s$ steps.
\end{thm}
\begin{pf}
	We can decompose the reduction from $\pi$ to its normal form $\pi_l$ as follows: $\pi=\pi_{-1}\red\pi_0\cdots\red\pi_l$, where each $\pi_i$ is $i$-normal.  By \reflemma{Linear}, if we call the length of the whole reduction sequence $L$, we have
	$$L\leq\sum_{i=0}^l\ISize{i}{\pi_{i-1}}\leq\sum_{i=0}^l\Size{\pi_{i-1}}.$$
	The reductions leading from $\pi_i$ to $\pi_{i+1}$ can be further decomposed as $\pi_i\red\pi_i'\red\pi_{i+1}$, where $\pi_i'$ is the first $i$-contractive \pn\ obtained in the reduction sequence.
	Observe now that the size of \pn s does not grow under non-contractive steps; therefore, for all $i$, $\Size{\pi_i'}\leq\Size{\pi_i}$. From this, if we apply \reflemma{SizeBoom}, we have that, for all $i$, $\Size{\pi_{i+1}}\leq2_2^{\Size{\pi_i}}$.

	It can now be proved by a straightforward induction that, for all $i\geq 0$, we have $\Size{\pi_{i-1}}\leq 2_{2i}^s$. Hence, we obtain
	$$L\leq\sum_{i=0}^l\Size{\pi_{i-1}}\leq\sum_{i=0}^l2_{2i}^s\leq(l+1)2_{2l}^s,$$
	as desired.\qed
\end{pf}

\begin{figure}[t]
	\begin{center}\scalebox{\scalefact}{\input{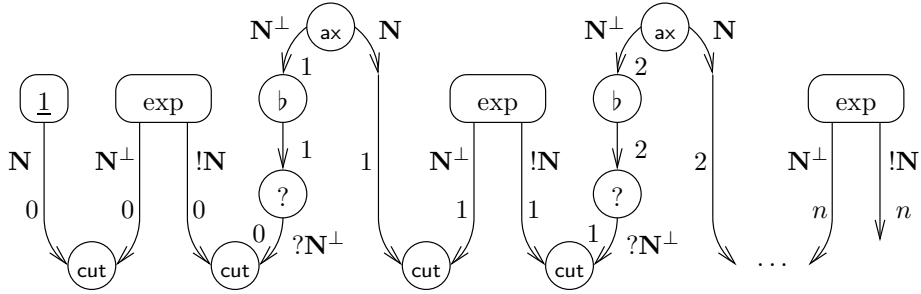}}\end{center}
	\caption{The \pn\ $\theta_n$, an iteration of $n$ \pn s computing the exponential function.}
	\label{fig:DepthLevelEx}
\end{figure}

Note that, in case we have a \MultELL\ \pn\ $\pi$ of size $s$ and depth $d$, by \refprop{ELL} depth and level coincide, so the above results tells us that $\pi$ can be reduced in at most $(d+1)2_{2d}^s$ steps, which is the bound found by \cite{DanosJoinet:ELL}. However, in \MELLlev\ it is in general the level that controls the complexity, not the depth. \reffig{DepthLevelEx} gives a clear example of this. It uses the fact that, following again \cite{DanosJoinet:ELL}, in \MultELL\ the exponential function $\mathrm{exp}(n)=2^n$ can be programmed as a \pn\ of conclusions $\Num^\bot,\oc\Num$, where $\Num$ is a suitable type of natural numbers, the cut-free \pn s of conclusion $\Num$ corresponding to Church integers, in analogy with the example given in \reffig{Exemplereseau}. Then, the cut-free form of the \pn\ $\theta_n$ of \reffig{DepthLevelEx} is the \pn\ representing the number $2_n$, i.e., a tower of powers of $2$ of height $n$. Hence, the size of $\theta_n$ is linear in $n$, but the size of its cut-free form is hyperexponential in $n$. This is in accordance with \refth{ElemBound}, because the level of $\theta_n$ turns out to be $n$. And yet, the depth of each $\theta_n$ is constant, indeed merely equal to $1$.

\subsection{Polynomial bound for \MLLLlev}
In the case of \MLLLlev, a finer analysis leads to a substantial improvement of \refth{ElemBound}. In the following, if a box $\cC$ contains a box $\cB$, we shall write $\cB\boxprec\cC$. The relation $\boxprec$ is obviously a finite, downward-arborescent partial order.
\begin{defn}[Light contractive order]
	Let $\pi$ be an \MELLlev\ \pn, and let $\cB,\cC$ be boxes of $\pi$. We put $\cB\immlghtctrprec\cC$ iff $\cB\immctrprec\cC$ and $\cB\not\boxprec\cC$. We denote by $\lghtctrprec$ the reflexive transitive closure of $\immlghtctrprec$, or, equivalently, we put $\cB\lghtctrprec\cC$ iff $\cB=\cC$, or $\cB\ctrprec\cC$ and $\cB\not\boxprec\cC$.
\end{defn}
\begin{lem}
	\label{lemma:Arborescence}
	In \MLLLlev, the relation $\lghtctrprec$ is an upward-arborescent partial order.
\end{lem}
\begin{pf}
	The fact that it is a partial order follows trivially from its definition and from \reflemma{POrd}, and indeed this is true for \MELLlev\ as well. For what concerns its arborescence, simply observe that, by the lightness condition of \refdef{MLLLlev}, for each box $\cC$ of an \MLLLlev\ \pn\ there may be at most one $\cB$ such that $\cB\immlghtctrprec\cC$.\qed
\end{pf}

\begin{figure}[t]
	\begin{center}\scalebox{0.65}{\input{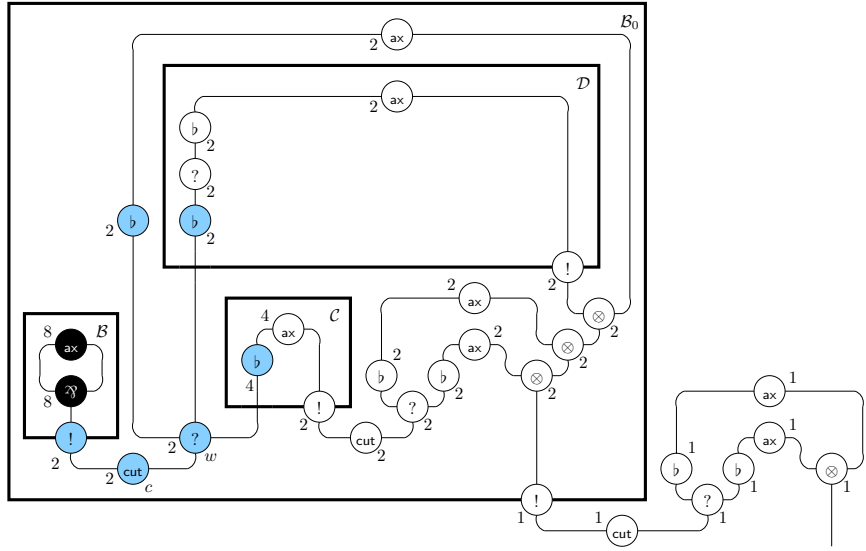}}\end{center}
	\caption{The \pn\ of \reffig{NestedLevel} (auxiliary ports are not drawn because irrelevant to the discussion of this section). Levels are omitted, since they are the same as those of \reffig{NestedLevel}. Instead, each link has its potential size relative to level~$0$ (see \refdef{PotSize}) annotated beside it.}
	\label{fig:RunningEx}
\end{figure}
Observe that, if $\cB,\cC$ are two boxes of an \MLLLlev\ \pn, thanks to the depth-stratification condition $\cB\immlghtctrprec\cC$ implies $\Depth{\cB}=\Depth\cC$. In fact, in \MLLLlev\ the light contractive order is simply a ``depth-wise slicing'' of the contractive order.

For example, if we take the \pn\ of \reffig{RunningEx}, we see that the contractive order at level $0$ is linear, i.e., $\cB\ctrprec\cC\ctrprec\cB_0$, while in the light contractive order we only have $\cB\lghtctrprec\cC$, and $\cB_0$ is incomparable with both $\cB$ and $\cC$, because it is not at the same depth.

\begin{defn}[Arity of a box]
	Let $\pi$ be an \MELLlev\ \pn, and let $\cB$ be a box of $\pi$. The \emph{arity} of $\cB$, denoted by $\Arity{\cB}$, is defined as follows:
	\begin{itemize}
		\item if the principal port of $\cB$ is premise of a \cut\ link whose other premise is the conclusion of a \whynot\ link $w$, then $\Arity{\cB}$ is equal to the arity of $w$ minus the number of \flatl\ links above $w$ which are inside a box $\cC$ such that $\cB\immlghtctrprec\cC$;
		\item otherwise, $\Arity{\cB}=1$.
	\end{itemize}
\end{defn}

Concretely, the arity of a box at level $i$ and depth $d$ is the number of copies that will be made of its content and that will not be subjected to further duplications by reducing cuts at level $i$ and depth $d$.

In the example of \reffig{RunningEx}, the \whynot\ link $w$ to which $\cB$ is cut has arity~$3$, but one of the \flatl\ links above it is inside a box $\cC$ such that $\cB\immlghtctrprec\cC$, hence $\Arity\cB=2$ (note that we do not have $\cB\immlghtctrprec\cD$ because $\cD$ is not at the same level as $\cB$). On the other hand, the arities of the other two boxes at level~$0$ are equal to the arities of their corresponding \whynot\ links: $\Arity\cC=2$ and $\Arity{\cB_0}=2$. Instead, since $\cD$ is not involved in a cut, $\Arity\cD=1$.
\begin{defn}[Contractive factor]
	Let $\pi$ be an \MELLlev\ \pn, and let $\cB$ be a box of $\pi$. The \emph{contractive factor} of $\cB$, denoted by $\CtrFact{\cB}$, is then defined as follows:
	$$\CtrFact{\cB}=\sum_{\cB\lghtctrprec\cC}\Arity{\cC}.$$
\end{defn}
\begin{lem}
	\label{lemma:CtrFact}
	Let $\pi$ be an \MLLLlev\ \pn, and let $\cB$ be a box of $\pi$. Then,
	$$\CtrFact{\cB}=\Arity{\cB}+\sum_{\cB\immlghtctrprec\cC}\CtrFact{\cC}.$$
\end{lem}
\begin{pf}
	Simply observe that, by \reflemma{Arborescence}, the set $\{\cC~;~\cB\lghtctrprec\cC\}$ can be partitioned into $\{\cB\}\cup\bigcup_{\cB\immlghtctrprec\cC}\{\cD~;~\cC\lghtctrprec\cD\}$.\qed
\end{pf}
\begin{defn}[Duplication factor]
	 Let $\pi$ be an \MELLlev\ \pn, and let $\cB$ be a box of $\pi$. The \emph{duplication factor} of $\cB$, denoted by $\Mult{\cB}$, is the following non-negative integer:
	$$\Mult{\cB}=\prod_{\cB\boxprec\cC}\CtrFact{\cC},$$
	where only boxes at the same level as $\cB$ are considered in the product.
\end{defn}

Still referring to \reffig{RunningEx}, we have $\CtrFact{\cB}=\Arity\cB+\Arity\cC=4$, while the contractive factors of $\cC$ and $\cB_0$ are equal to their arities, because these boxes are maximal in the light contractive order. This gives $\Mult\cB=\CtrFact\cB\CtrFact{\cB_0}=8$, $\Mult\cC=\CtrFact\cC\CtrFact{\cB_0}=4$, while $\cB_0$ is maximal w.r.t.\ $\boxprec$ and so \mbox{$\Mult{\cB_0}=\CtrFact{\cB_0}=2$}.

Intuitively, the duplication factor of a box $\cB$ at level $i$ says how many copies of the content of $\cB$ will be present at the end of the round at level $i$ of our cut-elimination procedure. In fact, the contractive factor takes into account the duplications originating from ``chains'' of boxes at the same depth; to obtain the duplication factor of a box $\cB$, one must multiply the contractive factors of all boxes containing $\cB$.

This is well shown in \reffig{RunningEx}: when one reduces the \cut\ link $c$, $3$ copies of the content of $\cB$ are made, but one of them will be duplicated again when the cut concerning $\cC$ is reduced, so $4=\CtrFact\cB$ copies are actually produced. We are not quite done though: the reduction of the cut concerning $\cB_0$ yields a further duplication of (the residues of) the content of $\cB$. Indeed, we invite the reader to check that exactly $8=\Mult\cB$ residues of the content of $\cB$ are present in the normal form of the \pn\ of \reffig{RunningEx}.

This motivates the following definition:
\begin{defn}[Potential size]
	\label{def:PotSize}
	Let $\pi$ be an \MELLlev\ \pn, and $k\in\Int$. The \emph{potential size} relative to $k$ of a link $a$ of $\pi$, denoted by $\PotSize{a}_k$, is defined as follows: let $\cB$ be the minimal box w.r.t.\ $\boxprec$ of level $k$ containing $a$; if $\cB$ exists, we set $\PotSize{a}_k=\Mult{\cB}$, otherwise $\PotSize{a}_k=1$. The \emph{potential size} relative to $k$ of $\pi$ is simply the sum of the potential sizes of its links:
	$$\PotSize{\pi}_k=\sum_a\PotSize{a}_k,$$
	where $a$ ranges over all links of $\pi$ which are not auxiliary ports.
\end{defn}

As suggested above, $\PotSize{\pi}_i$ is intended to give an estimate of the size of the \pn\ obtained by executing the round-by-round procedure at level $i$. This intuition is formalized by the following result:
\begin{lem}
	\label{lemma:PotSize}
	Let $\pi$ be an $i$-contractive \MLLLlev\ \pn. Then:
	\begin{enumerate}
		\item if $\pi$ is $i$-normal, then $\PotSize{\pi}_i=\Size{\pi}$;
		\item if $\pi\onered\pi'$ by reducing a minimal \cut\ link (in the cut order) at level $i$, then $\PotSize{\pi'}_i<\PotSize{\pi}_i$.
	\end{enumerate}
\end{lem}
\begin{pf}
	Part 1 is easy: simply observe that, if there is no reducible \cut\ link at level $i$, then for all $\cB$ at level $i$, by definition we have $\Arity{\cB}=1$. From this, since every box is maximal in the contractive order, we deduce $\CtrFact{\cB}=\Arity{\cB}=1$ for all $\cB$ at level $i$, and similarly $\Mult{\cB}=1$. This implies $\PotSize{a}_i=1$ for any link $a$ of $\pi$, which proves the result.

	The proof of part 2 is based on a careful inspection of \reffig{ExpStep}. We call the \whynot\ link and the box reduced by the step resp.\ $w$ and $\cB$. We also follow the convention that all links/boxes of $\pi$ will be denoted by ``simple'' letters ($a,\cC,\ldots$), while the links/boxes of $\pi'$ will be denoted by letters with a ``prime'' ($a',\cC',\ldots$); it shall be assumed that if the names of two links/boxes of resp.\ $\pi,\pi'$ differ only because of the absence/presence of a ``prime'', then one is the lift/residue of the other. For example, $a$ is the lift of $a'$, $\cC$ is the lift of $\cC'$, etc. The links of $\pi$ are partitioned into three classes (we ignore \pax s because they are not taken into account by the potential size):
	\begin{enumerate}
		\item[$C_1$:] links represented in \reffig{ExpStep} having a residue in $\pi'$; these are exactly the content of $\cB$ (i.e., the links contained in the \pps\ called $\pi_0$ in the picture), and, if present, the \whynot\ link of conclusion $\wn\Gamma$ (recall that, by the lightness condition, $\Gamma$ is at most one formula; if $\Gamma$ is empty, this link is not present);
		\item[$C_2$:] links represented in \reffig{ExpStep} having no residue in $\pi'$; these are exactly $w$, the \pal\ of $\cB$, the \cut\ link reduced by the step, and all of the \flatl\ links shown;
		\item[$C_3$:] all other links of $\pi$, i.e., those ``outside of the picture'' in \reffig{ExpStep}. These links have exactly one residue in $\pi'$.
	\end{enumerate}
	Similarly, the links of $\pi'$ can be partitioned into the following three classes:
	\begin{enumerate}
		\item[$C_1'$:] links having a lift of class 1 in $\pi$; these are exactly the links contained in one of the copies of $\pi_0$, and (if present) the \whynot\ link of conclusion $\wn\Gamma$;
		\item[$C_2'$:] links having no lift in $\pi$; these are exactly all of the \cut\ links represented in the right member of \reffig{ExpStep};
		\item[$C_3'$:] links having a lift of class 3 in $\pi$.
	\end{enumerate}
	The class of a box of $\pi$ or $\pi'$ will be the one of its \pal.

	Intuitively, in $\pi$ (resp.\ $\pi'$), a link of class 1 is a link which will be (resp.\ has been) duplicated or altered by the execution of the step; a link of class 2 is a link that disappears during (resp.\ is created by) the execution of the step; and a link of class 3 is a link to which ``nothing will happen'' (resp.\ ``nothing has happened'') during the execution of the step.

	\begin{figure}
		\begin{center}\scalebox{0.7}{\input{ex-running2.tex}}\end{center}
		\caption{The result of reducing the \cut\ link $c$ in the \pn\ of \reffig{RunningEx}.}
		\label{fig:RunningExRed}
	\end{figure}
	Before continuing with the proof, we invite the reader to pause a moment and look again at \reffig{RunningEx}. The \pn\ in the picture, which we denote by $\pi$, is readily seen to be $0$-contractive. As already noted above, the contractive order at level~$0$ is $\cB\ctrprec\cC\ctrprec\cB_0$, so the minimal cut in the cut order is the one denoted by $c$. After reducing it, we obtain the \pn\ $\pi'$ given in \reffig{RunningExRed}. In both figures, links filled with a dark shade are of class~1, those filled with a light shade are of class~2, and unfilled links are of class~3.

	We shall now verify part 2 of the lemma on this concrete example, by counting the links in each class and their potential sizes. We start with class~1 (dark-filled links). There are only $2$ such links in $\pi$: the \parl\ and \axiom\ link inside $\cB$. The deepest box of level $0$ containing them is precisely $\cB$, so their potential size is $\Mult\cB=8$. Therefore, the potential size of class~1 links of $\pi$ is $16$. For what concerns $\pi'$, we find $3$ copies of these two links: one inside $\cC'$, one inside $\cD'$, and one strictly inside $\cB_0'$. The first ones have potential size $\Mult{\cC'}=4$, and the last ones $\Mult{\cB_0'}=2$. For concerns the remaining copy, although it is contained in $\cD'$, this box has level $1$, so the potential size is again $\Mult{\cB_0'}=2$. Hence, the total potential size is $8+4+4=16$, i.e., identical to that of the links of class~$1$ of $\pi$.

	We may now turn to the links of class~$2$ (light-filled links). In $\pi$, there are $6$ of these, all of potential size $2$ except the \flatl\ link inside $\cC$, which has potential size $\Mult\cC=4$. The overall contribution to the potential size of $\pi$ from the links of class~$2$ is therefore $14$. In $\pi'$, all of these links have disappeared, and have been replaced by $3$ cuts at level $1$. Just as the \flatl\ links of class~$2$ in $\pi$, two of these \cut\ links have potential weight $2$, and one $4$, giving a total of $8<14$. Hence, in going from $\pi$ to $\pi'$ we have lost the potential size of the three links of class~$2$ of $\pi$ directly involved in the cut, i.e., the principal port of $\cB$, the \whynot\ link $w$, and the \cut\ link $c$ itself.

	Finally, we consider the links of class~$3$ (unfilled links). We invite the reader to check that, for each link $a$ of class~$3$ in $\pi$, there is exactly one residue $a'$ in $\pi'$, and $\PotSize{a}_0=\PotSize{a'}_0$. Therefore, the contribution to the potential size of the links in this class is preserved under reduction, and in the end we get $\PotSize{\pi'}_0<\PotSize{\pi}_0$, as stated in the lemma.

	We may now resume the proof. First of all, we recall the following fundamental fact, which holds by the  minimality of the cut under reduction:
	\begin{flushleft}
		\textbf{Fact}\quad\emph{If $\cB_1$ is a box of level $i$ such that $\cB_1\boxprec\cB$, then $\cB_1$ is not involved in a reducible cut}.
	\end{flushleft}
	The above fact can be used to infer the following series of preliminary results (before even reading the proofs, we strongly invite the reader to verify each one of them on the examples of Fig.~\ref{fig:RunningEx}~and~\ref{fig:RunningExRed}):
	\begin{claim}
		\label{claim:CtrOrd}
		Let $\cB_1',\cB_2'$ be two boxes of level $i$. Then, $\cB_1'\lghtctrprec\cB_2'$ iff $\cB_1\lghtctrprec\cB_2$.
	\end{claim}
	\begin{pf}
		Start by supposing that $\cB_1'\immlghtctrprec\cB_2'$. By definition, $\cB_1'$ is cut, by means of a \cut\ link $c'$, with a \whynot\ link above which there is exactly one (by the lightness condition) \flatl\ link inside $\cB_2'$. Observe that there are no \cut\ links of class~$1$ in $\pi'$, so $c'$ must be either of class~$2$~or~$3$. In the second case, obviously $\cB_1'$ and $\cB_2'$ are also of class~$3$3, so $\cB_1\lghtctrprec\cB_2$. The first case is actually impossible, because the premises of $c'$ would be of level $i+1$, hence none of them could be conclusion of the \pal\ of $\cB_1'$.

		Suppose now that $\cB_1\immlghtctrprec\cB_2$. Note firstly that we are supposing $\cB_1,\cB_2$ to be the lifts of resp.\ $\cB_1'$ and $\cB_2'$, so neither of $\cB_1,\cB_2$ can be equal to $\cB$. If they are both of class~$3$, we immediately have $\cB_1'\immlghtctrprec\cB_2'$. Suppose now that $\cB\immlghtctrprec\cB_1$. We cannot have $\cB\immlghtctrprec\cB_2$, because this would contradict \reflemma{Arborescence}. Therefore, $\cB_2$ is of class~$3$, and again obviously $\cB_1'\immlghtctrprec\cB_2'$. We are left with the case in which $\cB_1$ is of class~$1$ and $\cB\not\immlghtctrprec\cB_1$. The only possibility would be that $\cB_1\boxprec\cB$, but this is excluded by the above Fact, since we have supposed that $\cB_1$ is involved in a reducible cut. We have thus shown that $\cB_1'\immlghtctrprec\cB_2'$ iff $\cB_1\immlghtctrprec\cB_2$, which obviously implies our claim.\qed
	\end{pf}
	\begin{claim}
		\label{claim:Arity}
		Let $\cC'$ be a box of level $i$. Then, $\Arity{\cC'}=\Arity{\cC}$.
	\end{claim}
	\begin{pf}
		If $\cC'$ is not involved in a cut, then neither is $\cC$, so in this case the statement is obvious. In case $\cC'$ is involved in a cut $c'$, this cannot be one of the links of class~$2$ of $\pi'$, because they are all at level $i+1$. Therefore, $\cC$ is also involved in a cut, with a \whynot\ link that we may call $u$. Now notice that, if $u$ is of class~$3$, then the arities of $u$ and $u'$ coincide, and everything ``above'' $u$ is also of class~$3$, so the statement holds. But this is actually the only possibility: in fact, if $u$ were of class~$1$, it is easy to see that $u$ would have to be the unique (by the lightness condition) \whynot\ link such that, among its premises, there is (by the depth-stratification condition) the conclusion of the \pax\ of $\cB$. In this case, we would obtain $\cC\immctrprec\cB$, contradicting the minimality of the cut under reduction.\qed
	\end{pf}
	\begin{claim}
		\label{claim:NoNesting}
		If $\cB_1$ is a box of level $i$ such that $\cB_1\boxprec\cB$, then $\CtrFact{\cB_1}=1$.
	\end{claim}
	\begin{pf}
		In fact, $\CtrFact{\cB_1}>1$ would imply, by definition, that $\cB_1$ is involved in a contractive cut, which is impossible by the above Fact.\qed
	\end{pf}
	Claims~\ref{claim:CtrOrd}~and~\ref{claim:Arity} have the following fundamental corollary:
	\begin{claim}
		\label{claim:Mult}
		If $\cC$ is a box of class~$3$ of $\pi$ at level $i$, then $\Mult{\cC'}=\Mult{\cC}$.
	\end{claim}
	\begin{pf}
		Claims~\ref{claim:CtrOrd}~and~\ref{claim:Arity} immediately imply that, whenever $\cD$ is of class~$3$, $\CtrFact{\cD'}=\CtrFact{\cD}$. Now, any box containing a box of class~$3$ in $\pi$ is also of class~$3$, so if $\cD_1,\ldots,\cD_n$ are the nested boxes of level~$i$ surrounding $\cC$ in $\pi$, then in $\pi'$ we have boxes $\cD_1',\ldots,\cD'_n$ of level $i$ containing $\cC'$, with $\CtrFact{\cD'_j}=\CtrFact{\cD_j}$ for all $1\leq j\leq n$, which proves the claim.\qed
	\end{pf}

	Let now $a_3\in C_3$, and let $a_3'$ be its unique residue. It is not hard to see that, if $a_3$ is not contained in any box at level $i$, then neither is $a_3'$, in which case $\PotSize{a_3}_i=\PotSize{a_3'}_i=1$. Otherwise, let $\cB_0$ be the minimal box (w.r.t.\ $\boxprec$) of level $i$ containing $a_3$. Observe that $\cB_0\not\boxprec\cB$, because otherwise $a_3$ would not be of class~$3$. Therefore, $\cB_0$ has a unique residue $\cB_0'$, and both are of class~$3$. By \refclaim{Mult}, $\Mult{\cB_0}=\Mult{\cB_0'}$, so again $\PotSize{a_3}=\PotSize{a_3'}$. Recalling that every link of class~$3$ of $\pi$ has exactly one residue in $\pi'$, this shows that
	$$\sum_{a_3\in C_3}\PotSize{a_3}_i=\sum_{a_3'\in C_3'}\PotSize{a_3'}_i.$$

	Let instead $a_1\in C_1$. If $a_1$ is the \whynot\ link of conclusion $\wn\Gamma$, then it has a unique residue $a_1'$; in this case, by the same reasoning given above for links of class~$3$, we can easily infer that $\PotSize{a_1}_i=\PotSize{a_1'}_i$. Otherwise, $a_1$ is a link belonging to the \pps\ called $\pi_0$ in \reffig{ExpStep}. In this case, $a_1$ is contained in a box $\cB_1\boxprec\cB$ at level $i$; more precisely, there are $n$ boxes $\cB_1,\ldots,\cB_n$, all at level $i$, such that $a_1$ is in $\cB_1$ and $\cB_1\boxprec\cdots\boxprec\cB_n\boxprec\cB$, where each inclusion is immediate, i.e., there is no box at level $i$ between $\cB_j,\cB_{j+1}$ and $\cB_n,\cB$. Now, let $\Delta=\Mult{\cB_0}$, where $\cB_0$ is the minimal (w.r.t.\ $\boxprec$) box of level $i$ containing $\cB$, or let $\Delta=1$ if no such box exists. By \refclaim{NoNesting}, we have $\PotSize{a_1}_i=\Mult{\cB_1}=\Delta\CtrFact{\cB}$.

	Consider now a residue $a_2'$ of $a_2$. Each of the $\cB_j$ above has a corresponding residue $\cB_j'$ at level $i$ containing $a_2'$, such that $\cB_1'\boxprec\cdots\boxprec\cB_n'$. Since the structure of $\pi_0$ is not changed in the duplication, each $\cB_j'$ is maximal in the light contractive order and is not involved in a reducible cut, so $\CtrFact{\cB_j'}=1$ for all $j$. There are now two cases:
	\begin{enumerate}
		\item $\cB_n'$ is not contained in any box of level $i$, or the minimal (w.r.t.\ $\boxprec$) box containing it is $\cB_0'$. Then, $\PotSize{a_1'}_i=\Delta$. In fact, in case it exists, $\cB_0$ is of class~$3$, so by \refclaim{Mult}, $\Mult{\cB_0'}=\Mult{\cB_0}=\Delta$;
		\item There is a box $\cC'$ of level $i$ strictly contained in $\cB_0'$ and containing $\cB_n'$. In this case, by inspecting \reffig{ExpStep} under the depth-stratification condition, it is not hard to see that $\cC'$ is the unique residue of a box $\cC$ such that $\cB\immlghtctrprec\cC$. Observe that $\cC$ is of class~$3$, so by \refclaim{Mult} we have $\PotSize{a_1'}_i=\Mult{\cC'}=\Mult{\cC}=\Delta\CtrFact{\cC}$.
	\end{enumerate}
	If the arity of $w$ is $k\geq 1$, there are $k$ residues of $a_1$. Observe that case 1 applies to exactly $\Arity{\cB}$ of them, while case 2 applies to all other residues, and, because of the lightness condition, there is exactly one residue of this latter kind for each $\cC$ such that $\cB\immlghtctrprec\cC$. So, if we denote by $A_1'$ the set of all residues of $a_1$, we have, using \reflemma{CtrFact},
	$$\sum_{a_1'\in A_1'}\PotSize{a_1'}_i=\Delta\Arity{\cB}+\sum_{\cB\immlghtctrprec\cC}\Delta\CtrFact{\cC}=\Delta\CtrFact{\cB}=\PotSize{a_1}_i.$$
	If we put together what we have said up to now, we obtain an identical result for the links of class~$1$ as the one obtained above for the links of class~$3$:
	$$\sum_{a_1\in C_1}\PotSize{a_1}_i=\sum_{a_1'\in C_1'}\PotSize{a_1'}_i.$$

	We now get to the links of class~$2$, starting with those of $\pi$. The \pal\ of $\cB$, $w$, and $c$, have all potential size $\Delta$, where $\Delta$ is the same quantity introduced above. For what concerns the \flatl\ links shown in the picture, $\Arity{\cB}$ of them have again potential weight $\Delta$, while the others are each immediately (by the depth-stratification condition) contained in a different (by the lightness condition) box $\cC$ such that $\cB\immlghtctrprec\cC$, in which case the potential size is $\Delta\CtrFact{\cC}$. Therefore, we have
	$$\sum_{a_2\in C_2}\PotSize{a_2}_i=3\Delta+\Delta\Arity{\cB}+\sum_{\cB\immlghtctrprec\cC}\Delta\CtrFact{\cC}=\Delta(3+\CtrFact{\cB}).$$
	On the other hand, the only links of class~$2$ of $\pi'$ are the \cut\ links shown in the picture. Exactly $\Arity{\cB}$ of these have potential size $\Delta$, while the rest have each potential size $\Mult{\cC'}$, where $\cC$ is a box such that $\cB\immlghtctrprec\cC$ (of course we are implicitly using the above Claims to infer these facts). But, using \refclaim{Mult}, we have that $\Mult{\cC'}=\Mult\cC=\Delta\CtrFact{\cC}$, for all $\cC$ as above. Therefore, remembering that $\Delta\geq 1$, we obtain
	$$\sum_{a_2'\in C_2'}\PotSize{a_2'}_i=\Delta\Arity{\cB}+\sum_{\cB\immlghtctrprec\cC}\Delta\CtrFact{\cC}=\Delta\CtrFact{\cB}<\sum_{a_2\in C_2}\PotSize{a_2}_i,$$
	which concludes the proof of part 2.\qed
\end{pf}
We remark that the strict inequality of part~2 of \reflemma{PotSize} is a sort of an ``accident'', and is of no real technical value: what matters in the statement is that $\PotSize{\pi}_i$ linearly bounds $\PotSize{\pi'}_i$. \reflemma{SizeBound} below, which crucially uses \reflemma{PotSize}, would hold even if we only had $\PotSize{\pi'}_i=\PotSize{\pi}_i$, and indeed this is true at all levels except level $i$ itself, where the three links directly involved in the cut ``disappear'', and with them their potential size. More precisely, if we define the quantity $\PotSize{\pi}_i^j$ as the potential size relative to $i$ of all links of $\pi$ of level $j$, then point~2 of \reflemma{PotSize} can be replaced by $\PotSize{\pi'}_i^j=\PotSize{\pi}_i^j$ for all $i\neq j$ and $\PotSize{\pi'}_i^i<\PotSize{\pi}_i^i$.

As already noted above, the duplication factor of a box $\cB$ is influenced not only by the boxes $\cC$ at the same depth as $\cB$ such that $\cB\lghtctrprec\cC$, but also by the boxes at the same \emph{level} as $\cB$ which contain it. To quantify this phenomenon, we define the notion of \emph{relative depth}, which will be useful in bounding the potential size of a \pn\ (\reflemma{PotSizeBound}) and will be proved to have the same behavior as the level with respect to reduction, i.e., it is non-increasing (\reflemma{RelDepth}).
\begin{defn}[Relative depth]
	Let $\pi$ be an \MELLlev\ \pn, and let $\cB$ be a box of $\pi$. We denote by $\MBox{\cB}$ the maximal (w.r.t.\ $\boxprec$) box of $\pi$ at the same level as $\cB$ such that $\cB\boxprec\MBox{\cB}$. The \emph{relative depth} of $\cB$, denoted by $\RelDepth\cB$, is the following non-negative integer:
	$$\RelDepth{\cB}=\Depth{\cB}-\Depth{\MBox{\cB}}.$$
	The relative depth of $\pi$, also denoted by $\RelDepth{\pi}$, is the maximum relative depth of its boxes.
\end{defn}
Observe that, because $\boxprec$ is downward-arborescent, the relative depth of a box $\cB$ can be equivalently defined as the number of boxes $\cC$ at the same level as $\cB$ such that $\cB\boxprec\cC$, minus one.
\begin{lem}
	\label{lemma:PotSizeBound}
	Let $\pi$ be an \MELLlev\ \pn. Then, \mbox{$\PotSize{\pi}_i\leq{\Size\pi}^{\RelDepth\pi+2}$} for all \mbox{$i\in\Int$}.
\end{lem}
\begin{pf}
	Recall from the definition that $\PotSize{\pi}_i=\sum_a\PotSize{a}_i$, where the sum ranges over all links of $\pi$ other than \pax s. Now let $M=\max\{\PotSize{a}_i~;~a\in\pi\}$. Clearly we have that $\PotSize{\pi}_i\leq M\Size\pi$. Now $M$ must be the duplication factor of a box $\cB$ of level $i$ of $\pi$. For any such box, we have $\CtrFact{\cB}=\sum_{\cB\lghtctrprec\cC}\Arity{\cC}$. Observe that a \flatl\ link contributing to the arity of a box cannot contribute to the arity of another box; therefore, even if the sum defining $\CtrFact{\cB}$ ranged over \emph{every} box of $\pi$, we would still have $\CtrFact{\cB}\leq\Size{\pi}$. From this, recalling that the relative depth of a box $\cB$ of level $i$ is the number of boxes $\cC$ of level $i$ such that $\cB\boxprec\cC$, minus one, we have
	$$\Mult{\cB}=\prod_{\begin{array}{c}\scriptstyle\cB\boxprec\cC\\\scriptstyle\Level{\cC}=i\end{array}}\CtrFact{\cC}\leq\prod_{\begin{array}{c}\scriptstyle\cB\boxprec\cC\\\scriptstyle\Level{\cC}=i\end{array}}\Size\pi\leq{\Size\pi}^{\RelDepth\pi+1},$$
	which concludes the proof.\qed
\end{pf}
\begin{lem}
	\label{lemma:RelDepth}
	Let $\pi$ be an \MLLLlev\ \pn\ such that $\pi\onered\pi'$. Then, \mbox{$\RelDepth{\pi'}\leq\RelDepth\pi$}.
\end{lem}
\begin{pf}
	The depth of a box $\cC$ can only be affected during an exponential step, and only if it is contained in the \pps\ called $\pi_0$ in \reffig{ExpStep}. Then, if $\cC'$ is a residue of $\cC$ in $\pi'$, by the depth-stratification condition we either have $\Depth{\cC'}=\Depth{\cC}$ or $\Depth{\cC'}=\Depth{\cC}-1$, so in general $\Depth{\cC'}\leq\Depth\cC$.

	Now, call the box under reduction $\cB$; observe that $\cC\boxprec\cB$, so $\cB$ and $\MBox{\cC}$ cannot be disjoint. If we write $\cB_1\boxprecneq\cB_2$ for $\cB_1\boxprec\cB_2$ and $\cB_1\neq\cB_2$, then we can distinguish three cases: either $\MBox{\cC}\boxprecneq\cB$, or $\cB\boxprecneq\MBox{\cC}$, or $\MBox{\cC}=\cB$. In all cases, we put $\cD'=\MBox{\cC'}$.
	\begin{itemize}
		\item In the first case, the depth of $\cD'$ varies w.r.t.\ the depth of $\MBox{\cC}$ just as the depth of $\cC'$ varies w.r.t.\ the depth of $\cC$, so $\RelDepth{\cC'}=\RelDepth\cC$.
		\item 	In the second case, $\cD'$ is the unique residue of $\MBox{\cC}$, and $\Depth{\cD'}=\Depth{\MBox{\cC}}$, so $$\RelDepth{\cC'}=\Depth{\cC'}-\Depth{\cD'}\leq\Depth\cC-\Depth{\MBox{\cC}}=\RelDepth{\cC}.$$
		\item In the third case, we start by supposing that the lift $\cD$ of $\cD'$ is disjoint from $\cB$. Then, the depth-stratification condition gives us that $\cB\immlghtctrprec\cD$ and $\Depth{\cD'}=\Depth\cD=\Depth\cB$, so that $\RelDepth{\cC'}=\RelDepth\cC$. Suppose now that $\cD$ and $\cB$ are not disjoint. Since $\cB$ has no residue in $\pi'$, we have either $\cB\boxprecneq\cD$ or $\cD\boxprecneq\cB$. But the first case is actually impossible, because it would contradict the fact that $\cB=\MBox{\cC}$, since $\cD$ is at the same level as $\cC$. Therefore, we must have $\cD\boxprecneq\cB$, so that $\Depth{\cB}<\Depth{\cD}$. Now, as in the first case,
		$$\RelDepth{\cC'}=\Depth{\cC'}-\Depth{\cD'}=\Depth{\cC}-\Depth{\cD}<\Depth{\cC}-\Depth{\cB}=\RelDepth{\cC}.$$
	\end{itemize}\qed
\end{pf}

The technical machinery we have been building up through the section will now be used to finally infer our polynomial bound on the reduction of \MLLLlev\ \pn s.
\begin{lem}
	\label{lemma:SizeBound}
	Let $\pi$ be an $(i-1)$-normal \MLLLlev\ \pn, and let $\pi'$ be the $i$-normal \pn\ obtained from $\pi$ by applying the round-by-round procedure at level $i$. Then, $\Size{\pi'}\leq{\Size\pi}^{\RelDepth\pi+2}$.
\end{lem}
\begin{pf}
	We can decompose the reduction from $\pi$ to $\pi'$ into $\pi\red\pi_0\red\pi'$, where $\pi_0$ is the first $i$-contractive \pn\ obtained during the reduction. Now, applying, in the order, points 1~and~2 of \reflemma{PotSize}, \reflemma{PotSizeBound}, \reflemma{RelDepth}, and the well known fact that $\Size{\pi_0}\leq\Size\pi$, we obtain
	$$\Size{\pi'}=\PotSize{\pi'}_i\leq\PotSize{\pi_0}_i\leq{\Size{\pi_0}}^{\RelDepth{\pi_0}+2}\leq{\Size{\pi_0}}^{\RelDepth\pi+2}\leq{\Size\pi}^{\RelDepth\pi+2},$$
	as desired.\qed
\end{pf}
\begin{thm}[Polynomial bound for \MLLLlev]
	\label{th:PolyBound}
	Let $\pi$ be an \MLLLlev\ \pn\ of size $s$, level $l$, and relative depth $r$. Then, the round-by-round procedure reaches a normal form in at most $(l+1)s^{(r+2)^l}$ steps.
\end{thm}
\begin{pf}
	We start by applying the same arguments used in the beginning of the proof of \refth{ElemBound}: we decompose the reduction from $\pi$ to its normal form $\pi_l$ into $\pi=\pi_{-1}\red\pi_0\cdots\red\pi_l$, where each $\pi_i$ is $i$-normal; then, using \reflemma{Linear} (which is valid because \MLLLlev\ is a subsystem of \MELLlev), if we call the length of the whole reduction sequence $L$, we can write
	$$L\leq\sum_{i=0}^l\Size{\pi_{i-1}}.$$
	Now, using \reflemma{SizeBound}, we have, for all $0\leq i\leq l$, $\Size{\pi_i}\leq{\Size{\pi_{i-1}}}^{\RelDepth{\pi_{i-1}}+2}$. But, by \reflemma{RelDepth}, for all $0\leq i\leq l$, we have $\RelDepth{\pi_i}\leq\RelDepth\pi$, so we can actually write
	$$\Size{\pi_i}\leq{\Size{\pi_{i-1}}}^{r+2}.$$
	From this, it can be proved by a straightforward induction that, for all $i\geq 0$, we have $\Size{\pi_{i-1}}\leq s^{(r+2)^i}$. Hence, we obtain
	$$L\leq\sum_{i=0}^l\Size{\pi_{i-1}}\leq\sum_{i=0}^l s^{(r+2)^i}\leq(l+1)s^{(r+2)^l},$$
	which is the bound stated in the thesis.\qed
\end{pf}

Observe that, by \refprop{LLL}, if $\pi^+$ is the \MLLLlev\ embedding of an \MLLL\ \pn\ $\pi$ of size $s$ and depth $d$, then $\Size{\pi^+}=s$, $\Level{\pi^+}=d$, and $\RelDepth{\pi^+}=0$, so that normalizing $\pi^+$ takes at most $(d+1)s^{2^d}$ steps, which is the same bound given by \cite{Girard:LLL}.

\subsection{Characterization of \FE\ and \FP}
\label{sect:Characterization}

Propositions~\ref{prop:ELL}~and~\ref{prop:LLL} tell us that \MELLlev\ and \MLLLlev\ are conservative extensions of \MultELL\ and \MLLL, so programming in the former systems can be done using the same types and proofs as in the latter. In particular, the type of finite binary strings in \MELLlev\ and \MLLLlev\ are respectively
\begin{eqnarray*}
	\ElStr & = & \forall X.(\wn(X^\perp\ltens X)\lpar\wn(X^\perp\ltens X)\lpar\oc(X^\perp\lpar X)), \\
	\PolyStr & = & \forall X.(\wn(X^\perp\ltens X)\lpar\wn(X^\perp\ltens X)\lpar\S(X^\perp\lpar X)).
\end{eqnarray*}
Then, one can represent binary strings as in \cite{Girard:LLL} and \cite{DanosJoinet:ELL}. In the following, we write $\oc^k A$ (resp.\ $\S^k A$) for the formula $A$ preceded by $k$ of course (resp.\ paragraph) modalities, and if $\varphi$ and $\xi$ are two \pn s of respective conclusions $A^\perp,B$ and $A$, we denote by $\varphi(\xi)$ the \pn\ of conclusion $B$ obtained from $\varphi$ and $\xi$ by adding a \cut\ link whose premises are the conclusions of type $A^\perp,A$ of resp.\ $\varphi$ and $\xi$.
\begin{defn}[Representation]
	\label{def:Representation}
	A function $f:\{0,1\}^\ast\rightarrow\{0,1\}^\ast$ is \emph{representable} in \MELLlev\ (resp.\ \MLLLlev) iff there exists $k\in\Nat$ and a \pn\ $\varphi$ of conclusions $\ElStr^\perp,\oc^k\ElStr$ (resp.\ $\PolyStr^\perp,\S^k\PolyStr$) such that $f(x)=y$ iff $\varphi(\xi)\red\upsilon$, where $\xi$ is the \pn\ of conclusion $\ElStr$ (resp.\ $\PolyStr$) representing $x$, and $\upsilon$ is the \pn\ of conclusion $\oc^k\ElStr$ (resp.\ $\S^k\PolyStr$) which is the representation of $y$ enclosed in $k$ boxes (resp.\ followed by $k$ \pargl\ links). We denote by \FMELLlev\ (resp.\ \FMLLLlev) the class of functions representable in \MELLlev\ (resp.\ \MLLLlev).
\end{defn}

A fundamental remark now is that the level and relative depth of the representation of a datum do not depend on the datum itself: all cut-free \pn s of type $\ElStr$ representing binary strings in \MELLlev\ have level $1$, and all cut-free \pn s of type $\PolyStr$ representing binary strings in \MLLLlev\ have level $1$ and relative depth $0$. In both cases, the size of the \pn\ is equal to $3n+6$, where $n$ is the length of the string represented.

Thanks to the above, the soundness of \MELLlev\ and \MLLLlev\ with respect to \FE\ and \FP, respectively, is a consequence of Theorems~\ref{th:ElemBound}~and~\ref{th:PolyBound}, modulo the arguments given at the end of \refsect{ELLandLLL}. For the completeness side we have:
\begin{prop}
	\label{quanttativeCharacterization}
	 Any function $f:\{0,1\}^\ast\rightarrow\{0,1\}^\ast$ computable on a Turing machine in time $\mathcal O(2_d^n)$ can be represented in \MELLlev\ by a \pn\ of level $d$ and of conclusions $\ElStr^\perp,\oc^{d}\ElStr$.

	 Any function $f:\{0,1\}^\ast\rightarrow\{0,1\}^\ast$ computable on a Turing machine in time $\mathcal O(n^{2^d})$ can be represented in \MLLLlev\ by a \pn\ of level $d$ and of conclusions $\PolyStr^\perp,\S^{d}\PolyStr$.
\end{prop}
\begin{pf}
	Let us start with the second statement. First, \cite{MairsonTerui:CEP} show that a $\mathcal O(2_d^n)$ function can be represented in \MLLL\ by a \pn\ of depth $d$ and of conclusions $\PolyStr^\perp,\S^{d}\PolyStr$. Now we can obtain our statement by using the fact that any \MLLL\ \pn\ of depth $d$ gives an \MLLLlev\ \pn\ of level $d$ (\refprop{LLL}).

	As to the first statement, we have already recalled in the discussion after \refth{ElemBound} that \cite{DanosJoinet:ELL} give an encoding of the function $n \mapsto 2_d^n$ in \MultELL\ as a \pn\ of depth $d$ of conclusions $\Num^\bot,\oc^{d}\Num$, where $\Num$ is a type for tally integers. Using this fact and the encoding of Turing machines in \MultELL\ following the one from \cite{MairsonTerui:CEP}, we obtain that a function of $\mathcal O(2_d^n)$ can be represented in \MultELL\ by a \pn\ of depth $d$ and of conclusions $\ElStr^\perp,\oc^{d}\ElStr$. We then conclude as above, recalling that any \MultELL\ \pn\ of depth $d$ gives an \MELLlev\ \pn\ of level $d$ (\refprop{ELL}).\qed
\end{pf}

Hence, we finally have:
\begin{thm}[Characterization of \FE\ and \FP]
	\label{th:Characterization}
	\FMELLlev\ and \FMLLLlev\ coincide respectively with \FE\ and \FP.
\end{thm}
Observe that, due to the isomorphism $\parg(A\lpar B)\iso\parg A\lpar\parg B$, in \MLLLlev\ one may use the type $\PolyStr'=\forall X.(\wn(X^\perp\ltens X)\lpar\wn(X^\perp\ltens X)\lpar(\parg X^\perp\lpar\parg X))$ with virtually no difference, i.e., \refth{Characterization} still holds if we represent binary strings with this modified type.

\section{Restricting the Language of Formulas}
\label{sect:LLLlevz}
We have already observed that in \MLLLlev\ there are the following isomorphisms:
$$\parg(A\ltens B)\iso\parg A\ltens\parg B\qquad\parg\oc A\iso\oc\parg A\qquad\parg\forall X.A\iso\forall X.\parg A.$$
(Of course these isomorphisms hold in \MELLlev\ too, but we shall only deal with the polytime system in this section, since the paragraph modality is not really needed in \MELLlev). More generally, given a formula $A$ containing $\parg$, we may find several isomorphic formulas by commuting $\parg$ connectives with other connectives. This implies that given a proof $\pi$ of conclusion $A$, there are several computationally equivalent proofs that are obtained by composing $\pi$ with isomorphisms.

Hence, if we want to use \MLLLlev, or a fragment of it, as a type system for \lat s, we will have for each term
the choice between several types which carry essentially the same information.

A natural idea at this point is to choose a representative of each equivalence class of formulas, so as to obtain a ``canonical'' syntax. Given an \MLLLlev\ formula $A$, the obvious candidates to represent the equivalence class of $A$ are the formula in which all paragraphs have been pulled as close as possible to the root, and the formula in which all paragraphs have been pushed to the atoms. Clearly, only this latter choice is stable under composition of formulas (or prefixing with quantifiers and modalities); therefore, we shall draw our attention to the sublanguage of \MLLLlev\ in which \emph{$\parg$ connectives are only applied to atoms}, and we shall define a logical system, called \MLLLlevz, which uses such sublanguage.

To simplify the notations we shall replace $\S^p X$ by the notation $p X$ and let $p$ range over $\Nat$. Thus, the language of formulas of \MLLLlevz, denoted by $\canform$, will be generated by the following grammar:
$$A,B::=p X~|~pX^\perp~|~A\ltens B~|~A\lpar B~|~\oc A~|~\wn A~|~\exists X.A~|~\forall X.A,$$
where $p \in \Nat$. Linear negation is defined as expected: $(pX)^\perp=pX^\perp$, $(pX^\perp)^\perp=pX$, and $(\cdot)^\perp$ commutes with all connectives, replacing the given connective with its dual.

Given $p\in\Nat$ and a formula $A\in\canform$, we define $p\act A$ by induction on $A$ as follows:
\begin{eqnarray*}
	p \act (q X) &=& (p+q) X \\
	p \act (q X^\perp) &=& (p+q) X^\perp \\
	p \act (A \bullet B) &=& (p\act A)\bullet(p \act B), \mbox{ where } \bullet\in\{\ltens,\lpar\} \\
	p \act \mathop\dagger A &=& \mathop\dagger(p \act A), \mbox{ where } \dagger\in\{\oc,\wn\} \\
	p \act \nabla X.A &=& \nabla X.(p \act A), \mbox{ where } \nabla\in\{\forall,\exists\}. \\
\end{eqnarray*}
\begin{lem}
	\label{lemma:Action}
	For any $p, q \in \Nat$ and $A\in\canform$, we have
	\begin{eqnarray*}
		p\act (q\act A) & = & (p+q)\act A, \\
		0 \act A & = & A.
	\end{eqnarray*}
	Therefore, $\act$ is a monoid action on $\canform$.
\end{lem}
It is a straightforward consequence of the definition that whenever a formula $A\in\canform$ is equal to $p\act B$ for some $B$, then all subformulas of $A$ are also of the form $p\act B'$ for some subformula $B'$ of $B$. Also, it is easy to check that $(p \act A)^\perp=p\act A^\perp$.

In the language of formulas we could actually let $p$ range over $\Int$ instead of $\Nat$, and define a group action. We would then keep the same properties, but here we stick to $\Nat$ in order to have a clearer correspondence with \MLLLlev\ (that will be described below).

We now introduce a notion of substitution adapted to the formulas of $\canform$:
\begin{defn}
	For $A, B \in \canform$ we define $\cansubst{A}{B}{X}$ by induction on $A$:
	\begin{itemize}
		\item if $A=p X$: $\cansubst{p X}{B}{X}= p \act B$, 
		\item if $A=p X^\perp$: $\cansubst{p X^\perp }{B}{X}= p \act B^\perp$,
		\item and $\cansubst{}{B}{X}$ commutes with all connectives; for instance, $$\cansubst{(A_1\ltens A_2)}{B}{X}=\cansubst{A_1}{B}{X}\ltens\cansubst{A_2}{B}{X}.$$
	\end{itemize}
\end{defn}

We may now proceed to introducing the system \MLLLlevz. For this, we first need to define a suitable class of \pn s using the formulas of $\canform$.
\begin{defn}[\MELLz\ \pn s]
	\label{def:MELLz}
	The \ps s of \MELLz\ are defined as in \refdef{ProofStructure}, but for the following modifications (w.r.t.\ \reffig{Links}):
	\begin{itemize}
		\item edges are labelled by formulas in $\canform$;
		\item there is no \pargl\ link;
		\item \axiom\ links may have conclusions $p \act A^{\perp}$, $A$, for any $p\in\Nat$;
		\item \exist\ links have premise and conclusion with resp.\ types $\cansubst{A}{B}{X}$ and $\exists X.A$.
	\end{itemize}
	The \pn s of \MELLz\ are defined from these \ps s as in \refdef{ProofNet}.
\end{defn}

The intuition behind the unusual typing of the \axiom\ link is that it corresponds in \LLLlev\ to a proof of  $\parg^k A^{\perp}, A$, so an axiom followed by a series of \pargl\ links. However in \MLLLlevz\ paragraphs are only on atoms, and this is why we have a conclusion $p\act A^{\perp}$ instead of $\parg^k A^{\perp}$.

Cut-elimination for \MELLz\ \pn s is defined as in \MELL\ (Figures~\ref{fig:AxStep} through~\ref{fig:PargNoBoxStep}), except for the quantifier step (\reffig{QuantStep}), which uses the substitution $\cansubst{A}{B}{X}$ instead of $A[B/X]$, and for the axiom step (\reffig{AxStep}), which is treated as follows.

Let $\pi$ be a \MELLz\ \pn, and let $e$ be an edge of $\pi$. We say that a link $l$ of $\pi$ is \emph{above} $e$ if there exists a directed path from the conclusion of $l$ to $e$. We define the \emph{tree} of $e$, denoted by $\Tree e$, as the tree (ignoring boxes) whose root is $e$ and whose leaves are the conclusions of all the \axiom\ and \weak\ links above $e$. The \axiom\ links above $e$ are partitioned into three classes:
\begin{itemize}
	\item a \emph{neutral axiom} is an \axiom\ link such that both of its conclusions are leaves of $\Tree e$;
	\item a \emph{negative axiom} is an \axiom\ link whose conclusions are labelled by $p\act A^\perp$, $A$ and such that only the conclusion labelled by $p\act A^\perp$ is a leaf of $\Tree e$;
	\item a \emph{positive axiom} is an \axiom\ link whose conclusions are labelled by $p\act A^\perp$, $A$ and such that only the conclusion labelled by $A$ is a leaf of $\Tree e$.
\end{itemize}
If, in the negative or positive case, $p=0$, then the axiom may be considered as either positive or negative.

Now, suppose that $\pi$ contains a \cut\ link such that one premise is $e$ and the other premise is the conclusion $e'$ of an \axiom\ link $a$. The reduction of such a cut depends on whether $a$ is positive or negative with respect to $e'$ (it cannot be neutral, because $\Tree{e'}$ has only one leaf, $e'$ itself):
\begin{description}
	\item[negative:] we may assume that $e'$ is labelled by $p\act A^\perp$, so that $e$ is labelled by $p\act A$ and the other conclusion $e''$ of $a$ is labelled by $A$. In this case, $\pi$ reduces to the \pn\ $\pi'$ obtained as follows:
	\begin{itemize}
		\item remove $a$, and make $e$ coincide with $e''$;
		\item since $e$ is labelled by $p\act A$, all formulas labelling the edges of $\Tree e$ must be of the form $p\act B$ (cf.\ the remark after \reflemma{Action}); then, in $\pi'$ replace each $p\act B$ with $B$. It is easy to see that such a tree will have conclusion $A$;
		\item after this relabeling, if an axiom is neutral w.r.t.\ $e$, its conclusions will change from $p\act B,q\act p\act B^\perp$ to $B,q\act B^\perp$, so its residue is a valid axiom of \MELLz; if an axiom is positive or negative w.r.t.\ $e$, there is nothing to check because only one of its conclusions has been affected.
	\end{itemize}
	\item[positive:] we may assume that $e'$ is labelled by $A^\perp$, so that $e$ is labelled by $A$ and the other conclusion $e''$ of $a$ is labelled by $p\act A$. In this case, $\pi$ reduces to the \pn\ $\pi'$ obtained as follows:
	\begin{itemize}
		\item remove $a$, and make $e$ coincide with $e''$;
		\item for each formula $B$ labelling an edge of $\Tree e$, in $\pi'$ label the corresponding edge with $p\act B$; it is easy to see that such a tree will have conclusion $p\act A$;
		\item it is also easy to check that all axioms in $\pi'$ are still correctly labelled, just as in the negative case.
	\end{itemize}
\end{description}
\begin{defn}[Indexing]
	\label{def:Indexingz}
	An indexing $I$ for a \MELLz\ \pn\ is defined as in \refdef{indexing} but for the following modification: if $e$, $e'$ are the conclusions of an \axiom\ link with respective types $p \act A^{\perp}$ and $A$, then $I$ should satisfy $I(e')=I(e)+p$.
\end{defn}
\begin{defn}[\MLLLlevz]
	\label{def:MLLLlevz}
	The system \MLLLlevz\ is composed of all the \pn s of \MELLz\ admitting an indexing as in \refdef{Indexingz} and satisfying the (Weak) Depth-stratification and Lightness conditions of \refdef{MLLLlev}.
\end{defn}
It only takes a (tedious) case-by-case inspection to check that the above definition is sound, i.e., that \MLLLlevz\ is stable under cut-elimination.

Note that, because of the constraint on \axiom\ links (\refdef{Indexingz}), the possibility of assigning an indexing to a \MELLz\ \pn\ depends on the typing, in sharp contrast with the case of \MELL\ \pn s. Because of this, defining an untyped version of \MLLLlevz\ cannot be done as easily as for \MLLLlev\ (i.e., just forgetting the formulas).

A possible solution is the following. Consider a family of ``$p$-links'', with $p\in\Nat^\ast$, to be added to the usual links of untyped \MELL\ \pn s. The effect of a $p$-link is to ``change the level by $p$'', i.e., a $p$-link has one premise and one conclusion, whose levels must be resp.\ $i+p$ and $i$ (if typed, a $p$-link would have premise $A$ and conclusion $p\act A$). We add the restriction that the premise of a $p$-link must be the conclusion of an axiom link, and that each axiom has at most one $p$-link ``below''. Cut-elimination handles $p$-links by suitably adapting the axiom steps to an untyped framework. We shall not give any detail of this; the informal sketch we just gave is enough for our purposes.

Surprisingly, normalization fails in this system: there are untyped \MLLLlev\ \pn s whose reduction goes on forever. Perhaps this is not so strange after all: these $p$-links basically add the possibility of ``changing the level at will'', hence they completely break the fundamental invariant of \MELLlev\ and \MLLLlev\ \pn s (in fact, the level of an untyped \MLLLlevz\ \pn\ may increase under reduction).

The above discussion implies that it is impossible to adapt the arguments of \refth{PolyBound} to prove a complexity bound for \MLLLlevz. Nonetheless, in the rest of the section we shall argue that this system still characterizes deterministic polytime computation.

In what follows, we denote by $\form$ the set of \MELL\ formulas as defined in \refsect{Formulas}, i.e., including the paragraph modality. We shall now introduce two translations between our two systems:
$$
\begin{array}{ccc}
	\textrm{\MLLLlev} &\xrightarrow{\trzero{(\cdot)}} & \textrm{\MLLLlevz} \\
	\textrm{\MLLLlev} &\xleftarrow{\trone{(\cdot)}} & \textrm{\MLLLlevz}
\end{array}
$$
We first define them on formulas; this is done by induction on the argument formula:
\begin{eqnarray*}
	\trzero{X} & = & 0X \\
	\trzero{(X^\perp)} & = & 0X^\perp \\
	\trzero{(\parg A)} & = & 1\act\trzero{A}
\end{eqnarray*}
and $\trzero{(\cdot)}$ commutes with the other connectives, e.g.
\begin{eqnarray*}
	\trzero{(A \ltens B) } & = &  \trzero{A} \ltens \trzero{B}
\end{eqnarray*}
Similarly,
\begin{eqnarray*}
	\trone{(pX)} & = & \parg^p X\\
	\trone{(pX^\perp)} & = & \parg^p X^\perp
\end{eqnarray*}
and $\trone{(\cdot)}$ commutes with all connectives, e.g.
\begin{eqnarray*}
	\trone{(A \ltens B) } & = &  \trone{A} \ltens \trone{B}
\end{eqnarray*}
Observe that $\trzero{(\cdot)} \circ \trone{(\cdot)}$ is the identity on $\canform$, while $\trone{(\cdot)} \circ \trzero{(\cdot)}$ sends $A\in\form$ to the ``canonical'' representative of its equivalence class, i.e., the formula with all $\parg$ pushed to the atoms.

We shall now define how $\trzero{(\cdot)}$ and $\trone{(\cdot)}$ behave on proofs. Let $\pi$ be an \MLLLlev\ \pn. We say that a link $l$ is \emph{below} an edge $e$ or, equivalently, that $e$ is \emph{above} $l$ if in $\pi$ there is a directed path from $e$ to the premise of $l$. We then define $\trzero{\pi}$ as follows:
\begin{itemize}
	\item replace each axiom of conclusions $A^\perp, A$ by an axiom of conclusions \mbox{$q\act A^\perp$}, $p\act A$ where $q$ (resp.\ $p$) is the number of paragraph links below $A^\perp$ (resp.\ $A$) in $\pi$;
	\item remove paragraph links, and label each edge according to the relabeling of the axioms.
\end{itemize}
Informally speaking, $\trzero{\pi}$ is obtained from $\pi$ by pushing paragraph connectives upwards in the \pn, and ``absorbing'' them into the axioms. We have:
\begin{prop}
	\label{prop:TrZero}
	Let $\pi$ be an \MLLLlev\ \pn\ of conclusions $\Gamma$; then $\trzero{\pi}$ is an \MLLLlevz\ \pn\ of conclusions $\trzero{\Gamma}$.
\end{prop}
\begin{pf}
	Since $\pi$ is an \MLLLlev\ \pn\ it can be given an indexing $I$. To define an indexing $I_0$ on $\trzero{\pi}$ it is sufficient to define it on the conclusions of axioms. Each axiom link $a'$ in $\trzero{\pi}$ has conclusions $e_1',e_2'$ with respective types of the form $q\act A^\perp , p\act A$ and comes from an axiom $a$ of $\pi$ of conclusions $e_1,e_2$ with respective types $A^\perp$, $A$. W.l.o.g.\ we can assume $q\geq p$. Let $i=I(e_1)=I(e_2)$. Then set $I_0(e_1')=i-q$, $I_0(e_2')=i-p$. Note that we have $q \act A^\perp= (q-p)\act (p \act A)^\perp$ and $I_0(e_2')=I_0(e_1')+(q-p)$, so $I_0$ satisfies the condition on axioms, and is indeed an indexing. One can verify that $\trzero{\pi}$ is well-typed; a fundamental remark for this is that $\trzero{(\cdot)}$ preserves duality, i.e., $\trzero{(A^\perp)}=\trzero{A}^\perp$. To conclude, observe that the structure of $\pi$ and $\trzero{\pi}$ are basically identical: the only difference is the absence of paragraph links in $\trzero{\pi}$. But these are completely transparent to both the connected-acyclic condition (\refdef{ProofNet}) and the Depth-stratification and Lightness conditions (\refdef{MLLLlev}). Hence, since $\pi$ satisfies these conditions, so does $\trzero{\pi}$, which means that this latter is an \MLLLlevz\ \pn.\qed
\end{pf}

The translation $\trone{(\cdot)}$ requires a few preliminary definitions:
\begin{defn}
	Let $A\in\form$ and $p\in\Nat$; the \ps\ $R_A^p$ is defined as follows:
	\begin{itemize}
		\item let $S_A$ be the \MLLLlev\ \pn\ of conclusions $A^\perp, A$, representing the $\eta$-expansion of the axiom of conclusions $A^{\perp}, A$;
		\item $R_A^p$ is obtained from $S_A$ by replacing each axiom link of conclusion $X^\perp, X$, where $X^\perp$ is the type of the edge above the conclusion $A^\perp$, by the same link followed by $p$ paragraph links below $X^\perp$.
	\end{itemize}
\end{defn}
In the following, a \emph{weak} \MLLLlev\ \pn\ is a \MELL\ \pn\ satisfying the Depth-stratification and Lightness conditions (\refdef{MLLLlev}) and admitting a weak indexing.
\begin{lem}
	\label{lemma:PargEta}
	For all $A\in\form$ and $p\in\Nat$, $R_A^p$ is a weak \MLLLlev\ \pn.
\end{lem}
\begin{pf}
	A straightforward induction on $A$.\qed
\end{pf}

Let now $\pi$ be an \MLLLlevz\ \pn\ of conclusions $\Gamma$. Then, $\trone{\pi}$ is obtained by replacing each axiom of conclusions $p\act A^\perp,A$ in $\pi$ by $R_A^p$, and typing the rest of the edges accordingly.
\begin{prop}
	\label{prop:TrOne}
	Let $\pi$ be an \MLLLlevz\ \pn\ of conclusions $\Gamma$; then $\trone{\pi}$ is an \MLLLlev\ \pn\ of conclusions $\trone{\Gamma}$.
\end{prop}
\begin{pf}
	A more or less obvious corollary of \reflemma{PargEta}.\qed
\end{pf}

Observe that $\trzero{(\cdot)}\circ \trone{(\cdot)}$ does not act exactly as identity on \MLLLlevz\ \pn s, but performs an $\eta$-expansion. On the other hand, $\trone{(\cdot)}\circ \trzero{(\cdot)}$ behaves just like its counterpart on $\form$: given $\pi$, it gives the isomorphic \pn\ in which all paragraph links have been pushed to the axioms.

Both \MLLLlevz\ and \MLLLlev\ can be embedded in \MELL. For the first system, there is clearly a forgetful embedding $\Forget$ which simply erases the integers from atoms, both in formulas and proofs: $\Forget(pX)=X$, $\Forget(pX^\perp)=X^\perp$, and $\Forget$ commutes with all connectives. The second system is by definition a subsystem of \MELL, so the embedding would be trivial (the identity!); however, we are interested here in the following translation $\Erase{(.)}$:
\begin{itemize}
	\item given a formula $A\in\form$, $\Erase A$ is $A$ in which all $\parg$ have been removed;
	\item given an \MLLLlev\ \pn\ $\pi$, $\Erase\pi$ is $\pi$ in which all paragraph links have been removed, and types have been changed accordingly.
\end{itemize}
Clearly, both $\Forget$ and $\Erase{(.)}$ embed resp.\ \MLLLlevz\ and \MLLLlev\ in ``standard'' \MELL, i.e., multiplicative exponential linear logic \emph{without the paragraph modality} (actually, the embedding takes place in \MultELL). These two embeddings preserve cut-elimination:
\begin{lem}
	\label{lemma:Forget}
	Let $\pi$ be an \MLLLlevz\ \pn. Then, $\pi\onered\pi'$ iff $\Forget(\pi)\onered\Forget(\pi')$.
\end{lem}
\begin{pf}
	Simply observe that the untyped structure of $\pi$ and $\Forget(\pi)$ is identical, and cuts are reduced regardless of types (except quantifier cuts, but these are easily seen to be reciprocally simulated in one step).\qed
\end{pf}
\begin{lem}
	\label{lemma:Erase}
	Let $\pi$ be an \MLLLlev\ \pn. Then, $\pi\onered\pi'$ iff $\Erase\pi\red\Erase{(\pi')}$ in at most one step.
\end{lem}
\begin{pf}
	If $\pi\onered\pi'$, and the step applied is not a paragraph step, then clearly $\Erase\pi\onered\Erase{(\pi')}$. If it is a paragraph step, then it easy to see that $\Erase{(\pi')}=\Erase{\pi}$. For the converse, one reduction step in $\Erase\pi$ is always simulated by exactly one reduction step in $\pi$.\qed
\end{pf}
An important corollary of \reflemma{Forget} is the confluence and strong normalization of \MLLLlevz, which follows from the similar properties of \MELL\ \citep{Girard:LL}.

We also have a useful result relating the two embeddings:
\begin{lem}
	\label{lemma:Embeddings}
	Let $\pi$ be an \MLLLlev\ \pn. Then, $\Forget(\trzero{\pi})=\Erase\pi$.
\end{lem}
\begin{pf}
	As noted above, the translation $\trzero{(\cdot)}$ pushes paragraph links to the axioms, and then ``absorbs'' them into the formulas; then $\Forget$ forgets the annotations concerning paragraphs. But this amounts to simply removing the $\parg$ modality from both $\pi$ and its formulas.\qed
\end{pf}

\begin{figure}[t]
	\begin{center}\scalebox{\scalefact}{\input{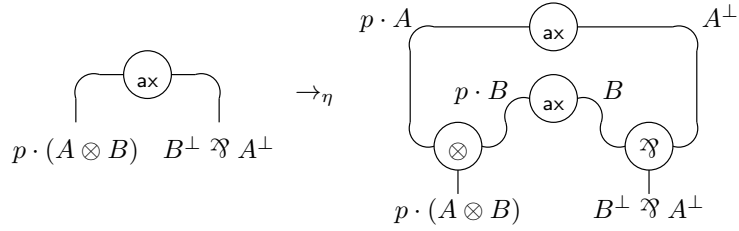}}\end{center}
	\caption{Multiplicative $\eta$-expansion step.}
	\label{fig:MultEtaStep}
\end{figure}
\begin{figure}[t]
	\begin{center}\scalebox{\scalefact}{\input{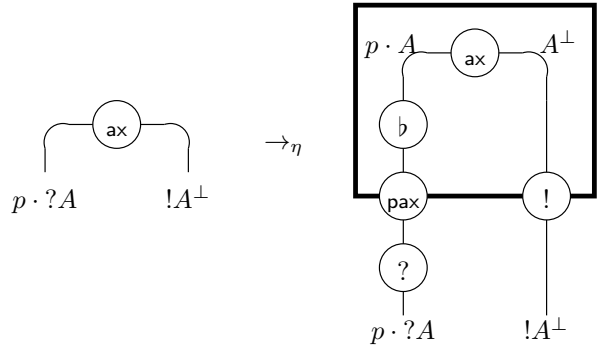}}\end{center}
	\caption{Exponential $\eta$-expansion step.}
	\label{fig:ExpEtaStep}
\end{figure}

In the sequel, we denote by $\oneetared$ the application of one $\eta$-expansion step to an \MLLLlevz\ \pn. One $\eta$-expansion step replaces a non-atomic axiom of conclusions $p\act C,C^\perp$ with axioms introducing the immediate subformulas of $C$. Figures~\ref{fig:MultEtaStep} and~\ref{fig:ExpEtaStep} give the definition for the cases $C=A\ltens B$ and $C=\wn A$; the other cases are treated similarly, as the reader may expect.
\begin{lem}
	\label{lemma:Eta}
	Let $\pi$ be an \MLLLlevz\ \pn\ such that $\pi\oneetared\pi_1\onered\pi_2$. Then, there exist $\pi_1',\pi_2'$ such that $\pi\onered\pi_1'\etared\pi_2''$ and $\pi_2'$ is $\beta$-equivalent to $\pi_2$, i.e., they have a common reduct through cut-elimination.
\end{lem}
\begin{pf}
	If the cut-elimination step applied in $\pi_1\onered\pi_2$ is ``far'' from the axioms, then the result is obvious. We can thus concentrate on the \emph{critical pairs}, i.e., the situations in which the axiom which is expanded in going from $\pi$ to $\pi_1$ is involved in a cut, and (the residue of) this cut is exactly the one reduced in going from $\pi_1$ to $\pi_2$. We check the only interesting case, leaving the others to the reader. Suppose that $\pi$ contains an axiom $a$ of conclusions $p\act\wn A,\oc A^\perp$, and the conclusion of type $\oc A^\perp$ is the premise of a cut $c$, whose other premise is the conclusion of a why not link $w$. We shall assume $p=0$; the general case is entirely similar. The $\eta$-expansion replaces $a$ with a box containing a \pps\ $\iota$ consisting of an axiom of conclusions $A,A^\perp$ and a flat link just below $A$. The cut-elimination step makes $n$ copies of $\iota$, and cuts them to the appropriate links. If we reduce these $n$ cuts, we obtain a \pn\ that we call $\pi_2'$. Now, if we take $\pi$ and reduce $c$ right away, it is immediate to see that we obtain exactly $\pi_2'$, and $\eta$-expansion is not even needed.\qed
\end{pf}

If $\pi$ is a \MELL\ or \MLLLlevz\ \pn, we denote by $\NF(\pi)$ its normal form, and by $\xrightarrow{\NF}$ reduction to the normal form. Then, we have:
\begin{lem}
	\label{lemma:BigDiag}
	The following diagrams commute:
	
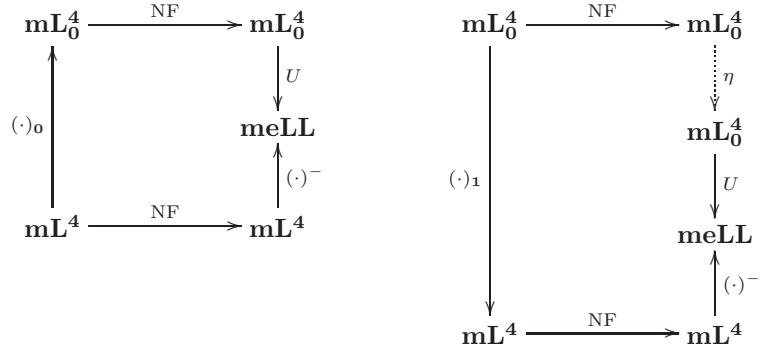
\begin{displaymath}
		\xymatrix{
			\mbox{\MLLLlevz} \ar[rr]^\NF && \mbox{\MLLLlevz} \ar[d]^\Forget \\
			& & \mbox{\MELL} \\
			\mbox{\MLLLlev} \ar[uu]^{\trzero{(\cdot)}} \ar[rr]^\NF && \mbox{\MLLLlev} \ar[u]_{\Erase{(\cdot)}}
		}
	\qquad\qquad
		\xymatrix{
			\mbox{\MLLLlevz} \ar[rr]^\NF \ar[ddd]_{\trone{(\cdot)}} & & \mbox{\MLLLlevz} \ar@{.>}[d]^\eta \\
			& & \mbox{\MLLLlevz} \ar[d]^\Forget \\
			& & \mbox{\MELL} \\
			\mbox{\MLLLlev} \ar[rr]^\NF & & \mbox{\MLLLlev} \ar[u]_{\Erase{(\cdot)}}
		}
	\end{displaymath}
	where the dotted arrow means that one may need to $\eta$-expand some axioms to close the second diagram.
\end{lem}
\begin{pf}
	For the first diagram, it is enough to prove that the three subdiagrams of the following diagram commute:
	
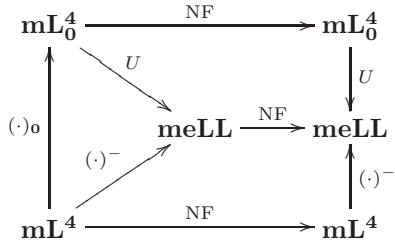
\begin{displaymath}
		\xymatrix{
			\mbox{\MLLLlevz} \ar[rr]^\NF \ar[dr]^\Forget && \mbox{\MLLLlevz} \ar[d]^\Forget \\
			& \mbox{\MELL} \ar[r]^\NF & \mbox{\MELL} \\
			\mbox{\MLLLlev} \ar[uu]^{\trzero{(\cdot)}} \ar[rr]^\NF \ar[ur]^{\Erase{(\cdot)}} && \mbox{\MLLLlev} \ar[u]_{\Erase{(\cdot)}}
		}
	\end{displaymath}
	These are consequences of Lemmas~\ref{lemma:Forget}, \ref{lemma:Erase}, and \ref{lemma:Embeddings}. For what concerns the second diagram, it is enough to prove that the three subdiagrams of the following diagram commute:
	
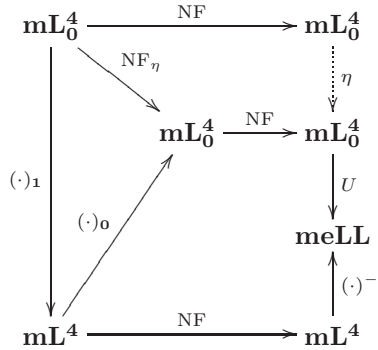
\begin{displaymath}
		\xymatrix{
			\mbox{\MLLLlevz} \ar[rr]^\NF \ar[ddd]_{\trone{(\cdot)}} \ar[dr]^{\NF_\eta} & & \mbox{\MLLLlevz} \ar@{.>}[d]^\eta \\
			& \mbox{\MLLLlevz} \ar[r]^\NF & \mbox{\MLLLlevz} \ar[d]^\Forget \\
			& & \mbox{\MELL} \\
			\mbox{\MLLLlev} \ar[uur]^{\trzero{(\cdot)}} \ar[rr]^\NF & & \mbox{\MLLLlev} \ar[u]_{\Erase{(\cdot)}}
		}
	\end{displaymath}
	where $\NF_\eta$ is the function associating with a \pn\ $\pi$ its $\eta$-expanded form, i.e., the \pn\ obtained by \mbox{$\eta$-expanding} all axioms of $\pi$ until only atomic axioms are left. Now, the commutation of the triangle on the left is simply the remark we made after \refprop{TrOne}, while the bottom subdiagram is nothing but the first diagram of this lemma.  Hence, all that is left to prove is the commutation of the top subdiagram. This is a consequence of \reflemma{Eta}. In fact, let $\pi$ be an \MLLLlevz\ \pn, and let $\pi'=\NF_\eta(\pi)$ and $\pi''=\NF(\pi')$. By definition, we have $\pi\etared\pi'\red\pi''$. We shall prove by induction on the length of the reduction $\pi\etared\pi'$ that $\NF(\pi)\etared\pi''$. If $\pi'=\pi$, then clearly $\NF(\pi)=\pi''$. If $\pi\etared\pi_1\oneetared\pi'$, then, using \reflemma{Eta}, by a further induction on the length of the reduction $\pi'\red\pi''$ we can prove that $\pi_1\red\pi_2\etared\pi_3$, and $\pi_3$ is \mbox{$\beta$-equivalent} to $\pi''$. But $\pi''$ is a normal form, so $\pi_2\etared\pi''$. Composing the reductions, we have $\pi\etared\pi_1\red\pi_2\etared\pi''$. Now the induction hypothesis applies, because the reduction $\pi\etared\pi_1$ is strictly shorter than $\pi\etared\pi'$. This gives us $\NF(\pi)\etared\pi_2\etared\pi''$, as desired.\qed
\end{pf}
Note that from the first diagram and \reflemma{Embeddings} we can infer that, for every \MLLLlev\ \pn\ $\pi$, $\Forget(\NF(\trzero{\pi}))=\Forget(\trzero{(\NF(\pi))})$. However, $\Forget$ is not injective, so we cannot conclude that the translation $\trzero{(\cdot)}$ commutes with reduction. The situation for the translation $\trone{(\cdot)}$ is even worse: $\Erase{(\NF(\trone{\pi}))}=\Erase{(\trone{(\NF(\pi))})}$ holds only up to $\eta$-equivalence.

We now proceed to argument how \MLLLlevz\ characterizes \FP\ (\refth{Characterizationz}). First of all, we define the \MLLLlevz\ type of finite binary strings as follows:
$$\Strz = \forall X.(\wn(0X^\perp\ltens 0X)\lpar\wn(0X^\perp\ltens 0X)\lpar(1X^\perp\lpar 1X)).$$
The reader can check that $\Strz=\trzero{(\PolyStr)}=\trzero{(\PolyStr')}$, where $\PolyStr$ and $\PolyStr'$ are the two isomorphic types that can be used for representing binary strings in \MLLLlev\ (cf.\ \refsect{Characterization}). Hence, by \refprop{TrZero}, if $\overline{x}$ is the \MLLLlev\ \pn\ of conclusion $\PolyStr$ (or $\PolyStr'$) representing the string $x$, the same string can be represented in \MLLLlevz\ by the \pn\ $\trzero{(\overline{x})}$.
\begin{lem}
	\label{lemma:Strings}
	Let $\xi,\xi'$ be two cut-free \pn s of resp.\ \MLLLlev\ and \MLLLlevz, of resp.\ conclusion $\parg^p\PolyStr$ (or $\trone{(\trzero{(\parg^p\PolyStr')})}$) and $p\act\Strz$, such that $\Forget(\xi')=\Erase\xi$. Then, $\xi$ and $\xi'$ represent the same binary string.
\end{lem}
\begin{pf}
	The fact that $\Forget(\xi')=\Erase\xi$ implies that $\xi$ and $\xi'$ have the same untyped structure modulo the presence of paragraph links in $\xi$; then the lemma is a consequence of the types of the two \pn s, and of the fact that they are cut-free.\qed
\end{pf}

Given a non-negative integer $p$ and an \MLLLlevz\ \pn\ $\pi$ not containing existential links, we denote by $p\act\pi$ the \pn\ obtained by replacing all atoms $A$ appearing in the types of $\pi$ with $p\act A$. It is easy to check that if $\pi$ is of conclusions $\Gamma$, then $p\act\pi$ is a well-typed \MLLLlevz\ \pn\ of conclusions $p\act\Gamma$. Moreover, if $\pi$ contains only atomic axioms, then so does $p\act\pi$.

In the following, if $\varphi$ is a \pn\ of conclusions $A^\perp,B$ and $\xi$ a \pn\ of conclusion $A$, we use the notation $\varphi(\xi)$ as introduced in \refsect{Characterization}. Observe that both $\trzero{(\cdot)}$ and $\trone{(\cdot)}$ are modular with respect to this notation, i.e., $\trzero{(\varphi(\xi))}=\trzero{\varphi}(\trzero{\xi})$ and $\trone{(\varphi(\xi))}=\trone{\varphi}(\trone{\xi})$.
\begin{defn}[Representation]
	Let $f:\{0,1\}^\ast\rightarrow\{0,1\}^\ast$. We say that $f$ is \emph{representable} in \MLLLlevz\ if there exists $p\in\Nat$ and an \MLLLlevz\ \pn\ $\varphi$ of conclusions $\Strz^\perp,p\act\Strz$ such that, whenever $\xi$ is a \pn\ of conclusion $\Strz$ representing the string $x$, we have $f(x)=y$ iff $\NF(\varphi(x))=p\act\upsilon$, where $\upsilon$ represents $y$.
\end{defn}
\begin{thm}
	\label{th:Characterizationz}
	Let $f:\{0,1\}^\ast\rightarrow\{0,1\}^\ast$. Then, $f\in\mbox{\FP}$ iff $f$ is representable in \MLLLlevz.
\end{thm}
\begin{pf}
	\sloppy{Let us start with the completeness of \MLLLlevz\ w.r.t.\ \FP. Let $f\in\mbox{\FP}$. By \refth{Characterization} there exist $p\in\Nat$ and an \MLLLlev\ \pn\ $\varphi$ such that, for all $x\in\{0,1\}^\ast$, $f(x)=y$ iff $\NF(\varphi(\xi))=\upsilon$, where $\upsilon$ is the representation of $y$ with $p$ paragraph links added to its conclusion. Let $\upsilon'=\NF(\trzero{(\varphi(\xi))})=\NF(\trzero{\varphi}(\trzero{\xi}))$. By the first diagram of \reflemma{BigDiag}, $\Erase\upsilon=\Forget(\upsilon')$, so by \reflemma{Strings} $\trzero{\varphi}$ represents~$f$.}

	For what concerns soundness, let $\varphi$ be an \MLLLlevz\ \pn\ of conclusions $\Strz^\perp,p\act\Strz$ representing the function $f$. For all $x\in\{0,1\}$, if $\xi$ is the \MLLLlevz\ representation of $x$, we have $f(x)=y$ iff $\NF(\varphi(\xi))=\upsilon'$, where $\upsilon'=p\act\upsilon$ and $\upsilon$ represents $y$. Now, observe that the representations of binary strings are all \mbox{$\eta$-expanded}, which means that $\upsilon'\etared\upsilon''$ implies $\upsilon''=\upsilon'$. Hence, in the second diagram of \reflemma{BigDiag} we can replace the dotted arrow with the identity, and obtain $\Forget(\upsilon')=\Erase{(\NF(\trone{(\varphi(\xi))}))}=\Erase{(\NF(\trone{\varphi}(\trone{\xi})))}$. The \pn\ $\NF(\trone{\varphi}(\trone{\xi}))$ is a normal form of type $\trone{(p\act\Strz)}=\trone{(\trzero{(\parg^p\PolyStr')})}$, so \reflemma{Strings} applies, and $\trone{\varphi}$ represents $f$ in \MLLLlev\ according to the alternative definition which uses the type $\PolyStr'$ for binary strings. But, as we pointed out in \refsect{Characterization}, \refth{Characterization} is still valid in this case, so $f\in\mbox{\FP}$.\qed
\end{pf}

\subsection{Sequent calculus for \MLLLlevz}
\label{sect:Seq0}
\begin{table}[t]
	\TwoRules
	{\infer[\etq{Axiom}]{\ts p\act A^{\perp i},A^{i+p}}{}}
	{\infer[\etq{Cut}]{\ts\Gamma,\Delta}{\ts\Gamma,A^i & \ts\Delta,A^{\perp i}}}
	\smallskip
	\TwoRules
	{\infer[\etq{Tensor}]{\ts\Gamma,\Delta,{A\ltens B}^i}{\ts\Gamma,A^i & \ts\Delta,B^i}}
	{\infer[\etq{Par}]{\ts\Gamma,{A\lpar B}^i}{\ts\Gamma,A^i,B^i}}
	\smallskip
	\begin{minipage}{\textwidth}
	\TwoRules
	{\infer[\etq{For all ($X$ not free in $\Gamma$)}]{\ts\Gamma,{\forall X.A}^i}{\ts\Gamma,A^i}}
	{\infer[\etq{Exists}]{\ts\Gamma,{\exists X.A}^i}{\ts\Gamma,{\cansubst{A}{B}{X}}^i}}
	\end{minipage}
	\smallskip
	\TwoRules
	{\infer[\etq{Light promotion}]{\ts\wn B^j,\oc A^i}{\ts B^{j+1},A^{i+1}}}
	{\infer[\etq{Dereliction}]{\ts\Gamma,\wn A^i}{\ts\Gamma,A^{i+1}}}
	\smallskip
	\TwoRules
	{\infer[\etq{Weakening}]{\ts\Gamma,\wn A^i}{\ts\Gamma}}
	{\infer[\etq{Contraction}]{\ts\Gamma,\wn A^i}{\ts\Gamma,\wn A^i,\wn A^i}}
	\caption{Rules for \MLLLlevz\ 2-sequent calculus. Daimon and mix are omitted.}
	\label{tab:MLLLlevzSeq}
\end{table}
It may be interesting to consider a sequent calculus formulation of \MLLLlevz, especially if one seeks to derive from it a type assignment system for the \lac, to be used to infer complexity properties about \lat s (in the style, for example, of \DLAL~\citep{BaillotTerui:DLAL}). Starting from the 2-sequent calculus for \MLLLlev\ (\refsect{Seq}), we end up with the rules given in \reftab{MLLLlevzSeq} (daimon and mix are again omitted, because identical to \reftab{MELLSeq}). As expected, weak \MLLLlevz\ \pn s correspond to derivations in this calculus, and \MLLLlevz\ \pn s to proper derivations. Observe the complete absence of a paragraph rule.

\section{Concluding Remarks and Further Work}
\label{sect:Conclusions}
We may perhaps summarize the fundamental contribution of the present work in one sentence: \emph{in linear-logical characterizations of complexity classes, exponential boxes and stratification levels are two different things}. From this fact, we have seen how one can define an elementary system extending \ELL, and a polynomial system extending \LLL. The main novelty of this latter, which is in direct connection with the above fact, is the absence of $\parg$-boxes. This implies that the paragraph modality commutes with all connectives; these commutations can be exploited to devise a polynomial system with a simpler class of formulas and fewer typing rules, which may be of interest for type assignment purposes. This is probably the most obvious direction of further research given by this work; in the sequel, we discuss other remarks and open questions.

\paragraph*{Indexes and tiers.} We already mentioned in the introduction how our form of stratification reminds of ramification, a technique devised by \cite{LeivantMarion93} to characterize complexity classes within the \lac. Ramification is enforced by so-called \emph{tiers}, which are integers assigned to subterms of a \lat, in close analogy with our indexes. However, we have not been able so far to understand the formal relationship between the two, and we suspect this may be an interesting subject for further work.

\paragraph*{Intensionality.}
\label{sect:Intensionality}
Concretely, the fact that \MELLlev\ and \MLLLlev\ extend resp.\ \MultELL\ and \MLLL\ means that the first two systems have ``more proofs'' that the latter two. Through the Curry-Howard looking glass, this means that \MELLlev\ and \MLLLlev\ are intensionally more expressive than Girard's corresponding systems, i.e., they admit ``more programs''. How many and which is still not clear though: we do have examples of \lat s  which are not typable in \MultELL\ and yet are typable in \MELLlev\ (or even in \MLLLlev!), but none of these corresponds to any ``interesting'' algorithm. So the question of whether our systems actually improve on the intensionality of \ELL\ and \LLL\ remains open.

\paragraph*{Naive set theory.}
\refprop{WN} states that, if we take an untyped \MELLlev\ \pn\ and start reducing its cuts, after a finite number of steps we either reach a cut-free form or a deadlock, i.e., a \pn\ whose all cuts are ill-formed. Now, the preservation of typing under reduction guarantees that, if the starting \pn\ is typed, then the latter case never happens; hence, \MELLlev\ satisfies cut-elimination.

This sharply contrasts with the situation one has in \MELL: weak normalization blatantly fails in untyped \MELL\ \pn s (the pure \lac\ can be translated in the system), and the proof of cut-elimination in the typed case is highly complex, because of the presence of second order quantification. Indeed, cut-elimination of second-order \MELL\ \pn s is known to be equivalent to the consistency of $\mathbf{PA}_2$~\citep{Girard:LL}, for which no inductive proof has ever been given (in other words, no-one knows what ordinal should replace $\omega^\omega$ in a proof like that of \refprop{WN}).

Following \cite{Girard:LLL} and \cite{Terui04}, one can build two naive set theories out of \MELLlev\ and \MLLLlev, which can still be proved to be consistent, i.e., to satisfy cut-elimination. In spite of their low logical complexity (as in the proof of \refprop{WN}, the consistency of these theories can be proved by an induction up to $\omega^\omega$), these set theories are particularly interesting because they are conservative extensions of the set theories based on elementary and light linear logic: they still use unrestricted comprehension, and thus allow arbitrary fixpoints of formulas, but they have more flexible logical principles, i.e., they admit more proofs. Asking how many more is of course another way of posing the above question about intensionality.

\paragraph*{Additives.}
\label{sect:Additives}
The additive connectives of linear logic ($\lwith$ and $\lplus$) have been excluded from this work; this is only a convenient choice, justified by the fact that some proofs (in particular those of \refprop{WN} and Theorems~\ref{th:ElemBound}~and~\ref{th:PolyBound}) become simpler. There is no technical problem in adding them to our systems, thus defining what we would call \LLlev\ and \LLLlev, which we still believe to exactly characterize resp.\ elementary and deterministic polytime computation.

There is however one point worth mentioning. The most natural definition of \LLLlev\ extends the commutation of the paragraph modality to additive connectives as well; in particular, the isomorphism $\parg(A\lplus B)\iso\parg A\lplus\parg B$ holds. \cite{Girard:LLL} has a nice argument against this being possible in \LLL, which goes as follows. For the sake of contradiction, suppose we can prove $\parg(A\lplus B)\llinimp\parg A\lplus\parg B$ in \LLL, and hence $\parg^p(A\lplus B)\llinimp\parg^p A\lplus\parg^p B$ for any $p\in\Nat$. Booleans can be easily encoded using the type $V_1\lplus V_2$, where $V_1$ and $V_2$ are two formulas admitting exactly one proof (for example
$V_1=V_2=\forall X.(X^\perp\lpar X)$). By similar definitions and arguments to those of \refdef{Representation} and
\refth{Characterization}, any language in $\mathbf{P}$ can be represented by an \LLL\ \pn\ $\varphi$ of conclusions
$\PolyStr^\perp,\parg^p(V_1\lplus V_2)$ for a suitable value of $p$ depending on the language itself. Now, using the commutation of the paragraph modality, we can transform $\varphi$ into a \pn\ $\varphi'$ of $\PolyStr^\perp,\parg^p V_1\lplus\parg^p V_2$. If we want to know whether the string $x$ belongs to our language or not, we
may simply take the \pn\ $\xi$ representing $x$ and normalize $\varphi'(\xi)$ (we are using the notation of
\refsect{Characterization}), which has conclusion $\parg^p V_1\lplus\parg^p V_2$. Observe that the main connective of this formula is $\oplus$, hence the $\mathsf{plus}$ link introducing it must be at depth zero, i.e., it is not contained in any exponential box. Observe also that the result of the computation is known as soon as the nature of this link is known, i.e., whether $\parg^p V_1\lplus\parg^p V_2$ is introduced from $\parg^p V_1$ or $\parg^p V_2$. But then, to have our answer, it is enough to stop the ``round-by-round'' cut-elimination procedure right after depth zero. In \LLL, normalizing just one depth is done in a number of steps linear in the size of the \pn, which can be done in quadratic time on a Turing machine, so we could solve any deterministic polytime problem in quadratic time, which is obviously false.

This argument however does not apply to \LLLlev\ because of the crucial difference between \emph{depth} and \emph{level}. A language in $\mathbf P$ may as well be represented in \LLLlev\ by a \pn\ $\varphi'$ of conclusions $\PolyStr^\perp,\parg^p V_1\lplus\parg^p V_2$, and it remains true that it is enough to normalize depth zero of
$\varphi'(\xi)$ to know whether the string represented by $\xi$ is in the language or not; however, the ``round-by-round'' cut-elimination procedure for \LLLlev\ goes \emph{level by level}, and depth zero may contain arbitrary many levels (in this case, $p$ levels is a good guess). Hence, normalizing just one depth may take a number of steps far from being linear in the size of the \pn, as we already showed in the example of \reffig{DepthLevelEx}.

\paragraph*{Denotational semantics.}
Recently, \cite{LaurentTortora:OCliques} proposed a denotational semantics for Girard's \ELL\ and Lafont's \SLL. Together with stratified coherence spaces~\citep{Baillot04a}, these are very interesting attempts at giving a completely semantic definition of complexity classes.

The present paper offers a new and arguably novel starting point in this perspective. With Boudes and Tortora de Falco, we are currently working on a categorical framework for building denotational semantics of \LLlev\ out of generic models of linear logic. From a syntactic point of view, this work is based on two alternative definitions of \LLlev, which do not make use of indices: the first one is \emph{geometric}, in the vein of correctness criteria; the second one is \emph{interactive}, i.e., it characterizes the \ps s of \LLlev\ in terms of their interactions with other \ps s.

\bibliographystyle{apalike}
\bibliography{LLlev}

\end{document}